\documentclass{aastex63}

\usepackage{tabularx}  
\usepackage{booktabs}  
\usepackage{amsmath}   
\usepackage{graphicx,subfigure}

\begin{document}

\title{GAMERA-OP: A three-dimensional finite-volume MHD solver for orthogonal curvilinear geometries}

\correspondingauthor{Binzheng Zhang}
\email{binzh@hku.hk}

\author{Hongyang Luo}
\affiliation{{Department of Earth and Planetary Sciences, the University of Hong Kong, Pokfulam, Hong Kong SAR}}
\affiliation{High Altitude Observatory, National Center for Atmospheric Research, United States of America}

\author{ Binzheng Zhang}
\affiliation{{Department of Earth and Planetary Sciences, the University of Hong Kong, Pokfulam, Hong Kong SAR}}
\affiliation{High Altitude Observatory, National Center for Atmospheric Research, United States of America}

\author{Jiaxing Tian}
\affiliation{{Department of Earth and Planetary Sciences, the University of Hong Kong, Pokfulam, Hong Kong SAR}}

\author{Jinshu Cai}
\affiliation{{Department of Earth and Planetary Sciences, the University of Hong Kong, Pokfulam, Hong Kong SAR}}

\author{Junjie Chen}
\affiliation{{Department of Earth and Planetary Sciences, the University of Hong Kong, Pokfulam, Hong Kong SAR}}

\author{Enhao Feng}
\affiliation{{School of Space and Earth Sciences, Beihang University, Beijing, China}}

\author{Zhiqi Zheng}
\affiliation{{Department of Earth and Planetary Sciences, the University of Hong Kong, Pokfulam, Hong Kong SAR}}

\author{Sheng Xi}
\affiliation{{Zhejiang Wellsun Intelligent Technology Co., Ltd., China}}

\author{John G. Lyon}
\affiliation{{Department of Physics and Astronomy, Dartmouth College}}



\begin{abstract}
We present GAMERA-OP (Orthogonal-Plus), a three-dimensional finite-volume magnetohydrodynamics (MHD) solver for orthogonal curvilinear geometries. The solver advances magnetic fields using constrained transport to preserve $\nabla\!\cdot\!\mathbf{B}=0$ to machine precision and employs geometry-consistent high-order reconstruction with an enhanced Partial Donor Cell method (e-PDM) that accounts for geometry curvature. Flexible numerics include various numerical fluxes and time integrators. In axial symmetric coordinates, angular momentum are preserved to round-off, and a ring-averaging treatment near the axis relaxes CFL constraints while maintaining divergence-free magnetic fields. Optional capabilities include the semi-relativistic (Boris) correction, background-field splitting, and an anisotropic MHD formulation. Rewritten in C, the code adopts a modular design that simplifies case setup and facilitates the addition of physics modules and coupling to other first-principles codes. Standard benchmarks across multiple geometries verify the code's high accuracy, low numerical diffusion, and robust handling of coordinate singularities and rotating flows. GAMERA-OP provides a practical, high-order framework for space and astrophysical plasma applications where orthogonal curvilinear coordinates and exact angular-momentum conservation are advantageous.

\end{abstract}


\section{Introduction}

Numerical simulations based on the magnetohydrodynamic (MHD) equations have long been a cornerstone for studying solar–terrestrial and astrophysical systems. Over recent decades, numerous well established MHD codes have been developed to  support astrophysical and space physics research, including ZEUS \citep{stone1992zeus}, Athena \citep{stone2008,stone2020}, PLUTO \citep{mignone2007}, NIRVANA \citep{ziegler2008nirvana}, the Pencil code \citep{dobler2006magnetic}, FLASH \citep{dubey2008introduction}, and BATS-R-US \citep{powell1999solution}. Adaptive mesh refinement (AMR) has increasingly been employed as an effective means to resolve phenomena with vastly different spatial and temporal scales \citep{stone2020,mignone2011pluto}, and high-order finite-volume methods have proven effective because they achieve higher accuracy at considerably lower cost relative to traditional second-order frameworks \citep{colella2011high,felker2018fourth}.

The Lyon–Fedder–Mobarry (LFM) MHD code, developed at the Naval Research Laboratory in the early 1980s, pioneered three-dimensional MHD in non-orthogonal curvilinear geometries using high-quality advection schemes. Its core design philosophy remains influential, employing high-order reconstruction with constrained transport (CT) \citep{lyon1981,evans1988} to enforce $\nabla\!\cdot\!\mathbf{B}=0$ to machine precision. LFM has been coupled to additional physics components, e.g., including a magnetosphere–ionosphere (M–I) coupler \citep{merkin2010effects} and the Rice Convection Model (RCM) for the inner magnetosphere \citep{pembroke2012initial}, and applied to problems ranging from the inner heliosphere \citep{merkin2011disruption} to multifluid space environment modeling \citep{brambles2011magnetosphere,lyon2025multifluid}. The GAMERA (Grid-Agnostic MHD for Extended Research Applications) code \citep{zhang2019} is a modern reinvention of LFM, retaining its strengths while significantly improving implementation, optimization, and numerical algorithms. GAMERA has been used extensively for the terrestrial magnetosphere \citep{sorathia2020ballooning}, the inner heliosphere \citep{provornikova2024mhd}, giant-planet magnetospheres \citep{zhang2021,zhang2024unified}, and unmagnetized bodies via a multifluid extension \citep{dang2023new}.

In LFM and GAMERA, the governing equations are solved on non-orthogonal curvilinear meshes using a ingeniously designed Cartesian-based finite-volume formulation. In this approach,  cell-centered vector components are expressed in a global Cartesian basis regardless of the computational coordinate system. While elegant and effective, this design complicates extension to AMR and to higher-order finite-volume frameworks. In addition, the Cartesian-based update does not strictly conserve angular momentum, which is essential for rapidly rotating systems. By contrast, solving the equations intrinsically in orthogonal coordinates simplifies higher-order FV implementation, and orthogonal curvilinear systems—such as cylindrical, spherical, and modified dipole \citep{kageyama2006note} —allow for exact angular momentum conservation.

Because ideal MHD simulation is a grid-scale fluid approximation, it cannot capture kinetic (particle-scale) physics; thus, coupling an MHD fluid solver to a subgrid or particle module is often desirable. However, the architecture of GAMERA makes such tight coupling, for example to particle solvers such as \citet{wang2017accurate}, comparatively inefficient. Motivated by these considerations, we introduce GAMERA-OP (Orthogonal-Plus), a successor to LFM/GAMERA rewritten from scratch in C (the predecessors were in Fortran). GAMERA-OP solves the MHD equations on arbitrary orthogonal curvilinear coordinates, adds extended physics modules, and incorporates improved numerical algorithms. It enforces solenoidal magnetic field condition via constrained transport on orthogonal geometry and conserves angular momentum to round-off using an angular-momentum-preserving discretization. A geometry-consistent, very high-order spatial reconstruction is employed, and extended MHD with anisotropic pressure is supported. GAMERA-OP further offers flexible time integration and flux options, including a second-order Adams–Bashforth predictor–corrector and a third-order SSP Runge–Kutta for time advancement, together with both a Boltzmann-type gas-kinetic flux and a Rusanov (local Lax–Friedrichs) flux for interface fluxes; optional semi-relativistic (Boris) correction and background-field splitting are available for strongly magnetized regimes.

The remainder of this paper is organized as follows. Section~\ref{sec:numerics} details the numerical methods. Section~\ref{sec:tests} presents verification tests across various geometries demonstrating its accuracy and robustness of the new solver. Section~\ref{sec:summary} summarizes the work.

\section{Numerical Schemes}
\label{sec:numerics}
 
\subsection{Governing Equations}

We solve the following set of single-fluid,  normalized ideal MHD equations:

\begin{equation}
    \frac{\partial \rho}{\partial t}=-\nabla \cdot(\rho \boldsymbol{u})  \label{continuity equaiton}
\end{equation}

\begin{equation}
    \frac{\partial \rho \boldsymbol{u}}{\partial t}=-\nabla \cdot(\rho \boldsymbol{u} \boldsymbol{u}+\overline{\boldsymbol{I}} P)+\boldsymbol{F_L} \label{momentum equation}
\end{equation}

\begin{equation}
    \frac{\partial \mathcal{E_P}}{\partial t}=-\nabla \cdot\left[\boldsymbol{u}\left(\mathcal{E_P}+P\right)\right]+\boldsymbol{u} \cdot \boldsymbol{F_L} \label{Ep equation}
\end{equation}

\begin{equation}
\frac{\partial \mathcal{E_T}}{\partial t}= -\nabla \cdot\left[ \boldsymbol{u} \left(\mathcal{E_T} + P\right) +\boldsymbol{E} \times \boldsymbol{B}\right]  \label{Etot equation}
\end{equation}

\begin{equation}
    \frac{\partial S}{\partial t}=-\nabla \cdot(S \boldsymbol{u})  \label{entropy equaiton}
\end{equation}

\begin{equation}
    \frac{\partial \boldsymbol{B}}{\partial t}=-\nabla \times \boldsymbol{{E}}, \label{faraday}
\end{equation}

where $\rho$, $\boldsymbol{u}$, $P$ are the plasma density, bulk velocity and thermal pressure respectively. $\boldsymbol{F_L}=\nabla \cdot\left(\boldsymbol{B B}-\overline{\boldsymbol{I}} \frac{B^2}{2}\right)$ represents the Lorentz force and $\boldsymbol{B}$ is the magnetic field. $\boldsymbol{{E}}=-\boldsymbol{u} \times \boldsymbol{B}$ is the electric field from the ideal Ohm's law. The quantities $\mathcal{E_P}$, $\mathcal{E_T}$ and $S$ denotes the plasma/hydrodynamic energy, total energy and entropy density defined as follows:
\begin{equation}
 \mathcal{E_P} = \frac{1}{2} \rho u^{2}+\frac{P}{\gamma-1},
\end{equation} 

\begin{equation}
 \mathcal{E_T}=\mathcal{E_P} + \frac{B^2}{2},
\end{equation} 

\begin{equation}
    S= P \rho^{1-\gamma },
\end{equation}

where $\gamma$ is the ratio of specific heat.

For most applications, as in LFM and GAMERA, the semi-conservative formulation based on the plasma energy equation ($\mathcal{E}_P$) is adopted as the default. This approach is especially robust in low-$\beta$ regimes, such as planetary magnetospheres and the solar corona, where the use of the total energy equation can easily lead to negative pressure due to strong magnetic dominance. Besides the total energy equation, the entropy equation is also included in GAMERA-OP to improve stability and robustness in certain regimes.

\subsection{Grid Definitions and Notations}

In this paper, we consider numerical schemes formulated in standard Cartesian, cylindrical, and spherical coordinate systems. The finite-volume computational cell labeled by indices $\left(i,j,k\right)$ is defined within the zone $[x_{1;i-\frac{1}{2}},x_{1;i+\frac{1}{2}} ] \times [ x_{2;j-\frac{1}{2}},x_{2;j+\frac{1}{2}} ] \times [ x_{3;k-\frac{1}{2}},x_{3;k+\frac{1}{2}}]$, where $\left(x_1, x_2, x_3\right)$ are the general orthogonal computational space coordinates equal to $(x, y, z)$, $(R, \phi, z)$ and $(r, \theta, \phi)$ for Cartesian, cylindrical, and spherical coordinates, respectively. Edge lengths, face areas, and cell volumes are determined using appropriate geometric scale factors. Table~\ref{grid_notation} provides a summary of the grid metric notations and their calculation for each coordinate system.

\begin{table}[h]

\caption{The Grid Definitions and Index Notations Used in GAMERA-OP}
\label{grid_notation}
\begin{tabular}{@{}cccccc@{}}
\toprule
\textbf{Grid Variable} & \textbf{Grid Location} & \textbf{Cartesian} & \textbf{Cylindrical} & \textbf{Spherical} & \textbf{Grid Index}\\ \midrule
   
$L_1$   & Cell Edges \emph{(1)} & $\Delta x$         & $\Delta R$     & $\Delta r$      & $i , j + \frac{1}{2}, k + \frac{1}{2}$ \\
$L_2$   & Cell Edges \emph{(2)}& $\Delta y$         & $R_{i+1 / 2}\Delta \phi$ & $r_{i+1 / 2}\Delta \theta $     & $i + \frac{1}{2}, j , k + \frac{1}{2}$\\
$L_3$    & Cell Edges \emph{(3)}& $\Delta z$         & $\Delta z$     & $r_{i+1 / 2}\sin \theta_{j+1 / 2} \Delta \phi$ & $i + \frac{1}{2}, j + \frac{1}{2}, k$ \\
$A_1$       & Cell Faces \emph{(1)} & $\Delta y \Delta z$                & $R \Delta \phi \Delta z$         & $-r_{i+1 / 2}^2 \Delta \cos \theta \Delta \phi $           & $i+\frac{1}{2}, j, k$ \\
$A_2$      & Cell Faces \emph{(2)}  & $\Delta z \Delta x$             & $\Delta R \Delta z$           & $\frac{1}{2}\Delta r^2 \sin \theta_{j+1 / 2} \Delta \phi$  & $i, j+\frac{1}{2}, k$ \\
$A_3$       & Cell Faces  \emph{(3)}& $\Delta x \Delta y$             & $\frac{1}{2}\Delta R^2 \Delta \phi$            & $\frac{1}{2}\Delta r^2  \Delta \theta$ & $i, j, k+\frac{1}{2}$ \\ 
$V$       & Cell Volumes  & $\Delta x \Delta y \Delta z$             & $\frac{1}{2}\Delta R^2 \Delta \phi \Delta z$            & $-\frac{1}{3} \Delta r^3 \Delta \cos \theta \Delta \phi $ & $i,j,k$ \\
\bottomrule
\end{tabular}

\smallskip
\begin{minipage}{1000pt}
{\footnotesize Note: Here the $\Delta$ symbol denotes the difference computed of the argument between its value at the upper and lower zone \\ interfaces, e.g., $\Delta r^3 = (r_{i+1/2}^3 - r_{i-1/2}^3)$, and $\Delta \cos \theta = \cos \theta_{j+1/2} - \cos \theta_{j-1/2}$.}
\end{minipage}
\end{table}

\begin{figure}[htb!]
	\noindent\includegraphics[width=\textwidth]{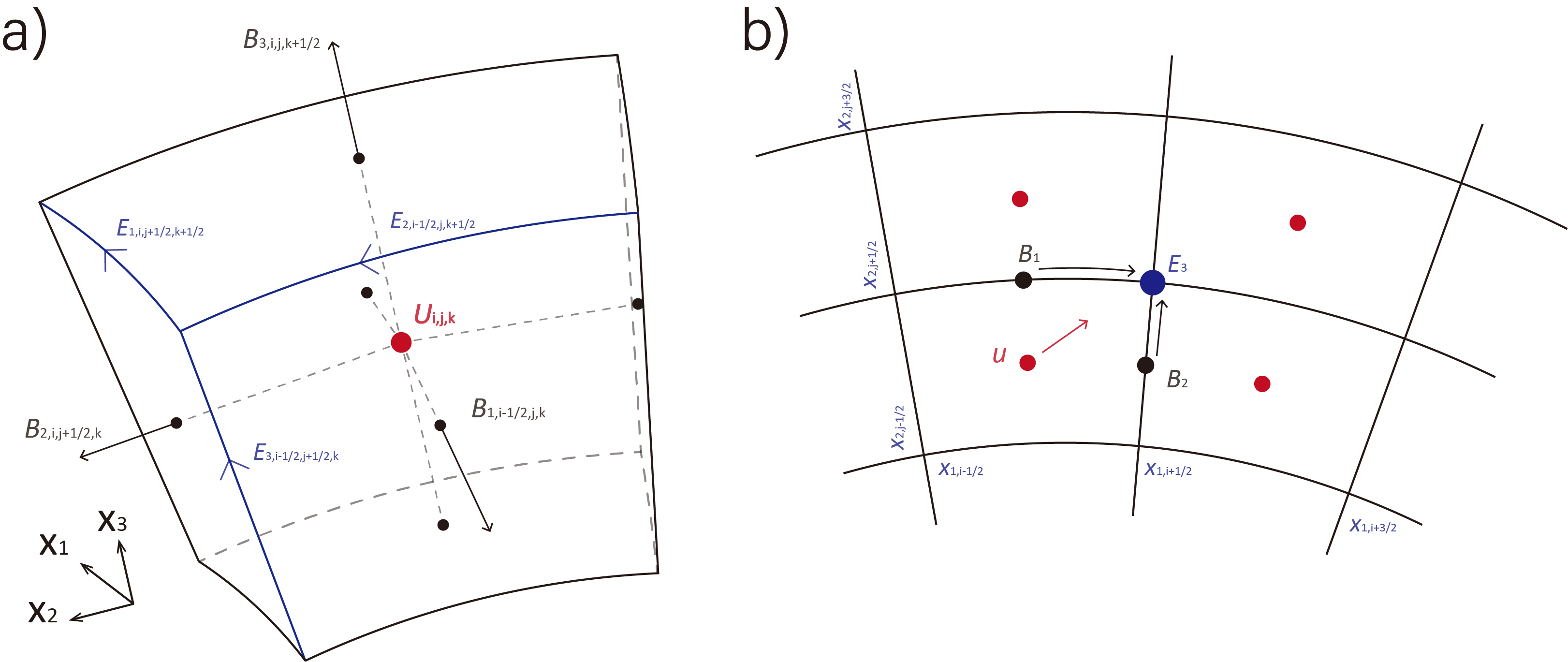}
	\centering
\caption{(a) The locations of the volume centered fluid variables $\boldsymbol{U}$, the face-centered magnetic fields $\boldsymbol{B}$, and the edge-centered electric fields $\boldsymbol{E}$. (b) A schematic showing the evaluation of the electric fields $E_3$ in a 2D slice ($x_1$-$x_2$ plane).}\label{grid plot}
\end{figure}

\subsection{Finite Volume Method}
\label{sec:FVM}
\subsubsection{The Fluid Subroutine}

Defining the conserved variable vector $U$:
\begin{equation}
    U = \{\rho,\rho \boldsymbol{u}, \mathcal{E_P},\mathcal{E_T},S \},
\end{equation}
the general finite-volume formulation of the governing equations~(\ref{continuity equaiton})–(\ref{entropy equaiton}) is obtained by integrating over each cell volume and applying Gauss’s law:

\begin{equation}
    \begin{aligned}
\frac{\partial}{\partial t} \int_V \boldsymbol{U} d V  =\oint \boldsymbol{F} \left(\boldsymbol{U},\boldsymbol{B}\right) \cdot d \boldsymbol{A}+\int_V \boldsymbol{S} d V,
\label{FVM_general}
\end{aligned}
\end{equation}

where $\boldsymbol{F}$ denotes the numerical fluxes through cell faces, and $\boldsymbol{S}$ represents source terms (both conservative and non-conservative), including geometric source terms arising from the tensor divergence in curvilinear coordinates.
The semi-discrete form of equation (\ref{FVM_general}) is given as follows:

\begin{equation}
    \begin{aligned}
\frac{\partial}{\partial t} \boldsymbol{U}_{i, j, k} = & \frac{1}{V_{i, j, k}} \Bigg[ \left(A_{1;i+\frac{1}{2}, j, k} \cdot {F}_{1;i+\frac{1}{2}, j, k} - A_{1;i-\frac{1}{2}, j, k} \cdot {F}_{1;i-\frac{1}{2}, j, k}\right) \\
& + \left(A_{2;i, j+\frac{1}{2}, k} \cdot {F}_{2;i, j+\frac{1}{2}, k} - A_{2;i, j-\frac{1}{2}, k} \cdot {F}_{2;i, j-\frac{1}{2}, k}\right) \\
& + \left(A_{3;i, j, k+\frac{1}{2}} \cdot {F}_{3;i, j, k+\frac{1}{2}} - A_{3;i, j, k-\frac{1}{2}} \cdot {F}_{3;i, j, k-\frac{1}{2}}\right) \Bigg]  + {S}_{i, j, k} \\
 \equiv & L_{\boldsymbol{U}} \left( \boldsymbol{U}, \boldsymbol{B}\right),
    \end{aligned}
\end{equation}

where $\boldsymbol{U}_{i, j, k}$ denotes the volume-averaged conserved variables and $L_{\boldsymbol{U}}$ represents the general RHS evaluation for the fluid variables. This expression is exact, and no approximation is made as long as $F$ and $S$ represent the proper face-averaged flux and volume-averaged source term respectively. In GAMERA-OP, we employ a formally second-order accurate finite-volume framework, wherein volume-, face-, and line-averaged values are used interchangeably with centered values. The framwork of GAMERA-OP allows further extensions of the FV solver to a formal order higher than two, which will be the focus of future development. 

\subsubsection{The Maxwell Subroutine}

GAMERA-OP employs the Yee-Grid \citep{yee1966} to enforce the solenoidal constraint $\nabla \cdot \boldsymbol{B} = 0$. With such a staggered grid, the primary magnetic fields are located at cell faces, and the electric field at cell edges are interpreted as the numerical fluxes of magnetic field integration. Applying Stokes’ theorem to the integral form of Faraday’s law (\ref{faraday}) yields:

\begin{equation}
    \int \frac{\partial \boldsymbol{B}}{\partial t} \cdot d \boldsymbol{A} =- \int \boldsymbol{{E}} \cdot d \boldsymbol{l}  
\end{equation}

As an result, the semi-discrete evolution equation for the magnetic field in the $x_1$-direction is

\begin{equation}
    \begin{aligned}
\frac{\partial}{\partial t} B_{1;i+\frac{1}{2}, j, k}= & -\frac{1}{{A_{1;i+\frac{1}{2},j,k}}}\left({E}_{2;i+\frac{1}{2}, j, k-\frac{1}{2}} \cdot L_{2;i+\frac{1}{2}, j, k-\frac{1}{2}}+{E}_{3;i+\frac{1}{2}, j+\frac{1}{2}, k} \cdot L_{3;i+\frac{1}{2}, j+\frac{1}{2}, k}\right. \\
& \left.-{E}_{2;i+\frac{1}{2}, j, k+\frac{1}{2}} \cdot L_{2;i+\frac{1}{2}, j, k+\frac{1}{2}}-{E}_{3;i+\frac{1}{2}, j-\frac{1}{2}, k} \cdot L_{3;i+\frac{1}{2}, j-\frac{1}{2}, k} \right)\\
\equiv & L_{\boldsymbol{B}} \left( \boldsymbol{U}, \boldsymbol{B}\right),
\end{aligned}
\end{equation}

where $B_{1;i+\frac{1}{2},j,k}$ is the face-averaged magnetic field and $E$ denotes the edge-averaged electric field and $L_{\boldsymbol{B}} $ represents the general RHS evaluation for the magnetic fields.

The detailed calculation of the electric fields is described in Section \ref{CT}. Summing the updates over all cell faces yields exact pairwise cancellation; therefore, the scheme conserves magnetic flux. The constrained transport method \citep{lyon1981,evans1988} preserves the divergence-free condition $\nabla \cdot \boldsymbol{B} = 0$ to round-off, provided it is initially satisfied.

Once face-centered magnetic fields are advanced, cell-centered values are required to  evaluate magnetic contributions to the numerical fluxes. These magnetic fields are obtained via linear interpolation from the face values to volumetric centers:
 
\begin{equation}
    B_{1,i}=\frac{x_{1,i+\frac{1}{2}}-\langle x_1 \rangle_i}{x_{1,i+\frac{1}{2}}-x_{1,i-\frac{1}{2}}}  B_{1,i-\frac{1}{2}}+\frac{\langle x_1 \rangle_i-x_{1,i-\frac{1}{2}}}{x_{1,i+\frac{1}{2}}-x_{1,i-\frac{1}{2}}}  B_{1,i+\frac{1}{2}},
\end{equation}

where $\langle x_1 \rangle_i$ represents the volumetric center (barycenter) coordinate,
\begin{equation}
    \langle x_1 \rangle_i =\frac{1}{V_{i j k}} \int_{V_{i j k}} x_1 d V .
\end{equation}

\begin{table}[h]
\centering
\caption{Volume-Averaged Coordinate}
\label{table barycenter}
\begin{tabular}{@{}cccc@{}}
\toprule
\textbf{Coordinate}  & \textbf{Cartesian} & \textbf{Cylindrical} & \textbf{Spherical} \\ \midrule
   
$\langle x_1 \rangle_i$   & $\frac{1}{2} ({x_{i+\frac{1}{2}}+x_{i-\frac{1}{2}}}) $ & $\frac{2}{3}\frac{\Delta R^3}{\Delta R^2}$         & $\frac{3}{4}\frac{\Delta r^4}{\Delta r^3}$          \\
$\langle x_2 \rangle_j$   & $\frac{1}{2}({y_{j+\frac{1}{2}}+y_{j-\frac{1}{2}}})$        & $\frac{1}{2}({\phi_{j+\frac{1}{2}}+\phi_{j-\frac{1}{2}}})$     & $\frac{\Delta(\theta \cos \theta)-\Delta(\sin \theta)}{\Delta (\cos \theta)}$  \\
$\langle x_3 \rangle_k$    & $\frac{1}{2}({z_{k+\frac{1}{2}}+z_{k-\frac{1}{2}}})$      & $\frac{1}{2}({z_{k+\frac{1}{2}}+z_{k-\frac{1}{2}}})$    & $\frac{1}{2}({\phi_{k+\frac{1}{2}}+\phi_{k-\frac{1}{2}}})$  \\
\bottomrule
\end{tabular}
\end{table}

The coordinates of the volume centroid are summarized in Table \ref{table barycenter}. Unlike LFM~\citep{lyon2004} and GAMERA~\citep{zhang2019}, which advance interface-threading magnetic fluxes, GAMERA-OP directly advances face-centered magnetic fields, a choice that streamlines implementation for orthogonal coordinate systems.

\subsection{Time Integration}
\label{time marching}
This section provides an overview of the time integration algorithms implemented in GAMERA-OP for advancing the MHD equations (\ref{continuity equaiton})--(\ref{faraday}). Two explicit time-marching schemes are available: a second-order Adams–Bashforth (AB2) predictor–corrector integrator and a three-stage, third-order strong-stability-preserving Runge–Kutta (SSP-RK3) scheme. Details of each method are described below.

\subsubsection{Second-order Adams–Bashforth (AB2) Scheme}
\label{AB2}

One of the available time integration methods in GAMERA-OP, following the approach of LFM~\citep{lyon2004} and GAMERA~\citep{zhang2019}, is the second-order Adams–Bashforth (AB2) scheme. The AB2 algorithm advances the solution through a predictor step, which extrapolates linearly to the temporal midpoint, followed by a corrector step that performs the full flux calculation using the predicted variables.

The predictor step for the conserved variables and the magnetic field is given by:

\begin{equation}
    \boldsymbol{U}^{n+\frac{1}{2}}=\boldsymbol{U}^n+\frac{\Delta t^n}{2 \Delta t^{n-1}}\left(\boldsymbol{U}^n-\boldsymbol{U}^{n-1}\right),
\end{equation}

\begin{equation}
    \boldsymbol{B}^{n+\frac{1}{2}}=\boldsymbol{B}^n+\frac{\Delta t^n}{2 \Delta t^{n-1}}\left(\boldsymbol{B}^n-\boldsymbol{B}^{n-1}\right).
\end{equation}

After computing the half-step values, the corrector step updates the solution using the temporal midpoint variables:

\begin{equation}
    \boldsymbol{U}^{ n+1} = \boldsymbol{U}^{ n}+\Delta \boldsymbol{U}^{n+\frac{1}{2}},
    \label{delta F M}
\end{equation}

where $\Delta \boldsymbol{U}^{n+\frac{1}{2}} = \Delta t \cdot L_{\boldsymbol{U}}\left(\boldsymbol{U}^{n+\frac{1}{2}}, \boldsymbol{B}^{n+\frac{1}{2}}\right)$, and $L_{\boldsymbol{U}}$ denotes the right-hand side of the finite-volume update (see Section~\ref{sec:FVM} ).

The magnetic field is advanced similarly:
 \begin{equation}
     \boldsymbol{B}^{n+1}=\boldsymbol{B}^{n}+\Delta \boldsymbol{B}^{n+\frac{1}{2}},
 \end{equation}
where $\Delta \boldsymbol{B}^{n+\frac{1}{2}} = \Delta t \cdot L_{\boldsymbol{B}}\left(\boldsymbol{U}^{n+\frac{1}{2}}, \boldsymbol{B}^{n+\frac{1}{2}}\right)$.

The time step $\Delta t$ is prescribed by the Courant–Friedrichs–Levy (CFL) condition for explicit schemes:

\begin{equation}
    \Delta t=N_{\mathrm{CFL}} \cdot \text{min}\left(\frac{\text{min} \left( h_1 \Delta x_1, h_2 \Delta x_2 , h_3 \Delta x_3  \right)}{v}\right)
\end{equation}

\begin{equation}
    v=|u|+\sqrt{V_A^2+C_S^2}
\end{equation}

where $|u|$ is the magnitude of the plasma bulk velocity and $\sqrt{V_A^2+C_S^2}$ is the magnetosonic speed estimated using cell-centered values. In practice, $N_{\mathrm{CFL}} = 0.3$ is typically used with the AB2 time integrator.

\subsubsection{Third-order Runge–Kutta}
\label{RK3}

GAMERA-OP provides a three-stage, third-order strong stability preserving Runge–Kutta (SSPRK3) method~\citep{gottlieb2009} for time advancement. The SSPRK3 scheme proceeds as follows for the conserved variables:

For the fluid subroutine:

\begin{equation}
    \boldsymbol{U}^{ *} = \boldsymbol{U}^{ n}+\Delta \boldsymbol{U}^n\\
\end{equation}

\begin{equation}
    \boldsymbol{U}^{**} = \frac{1}{4} \left(3\boldsymbol{U}^{ n}+\boldsymbol{U}^{ *}+\Delta \boldsymbol{U}^* \right)
\end{equation}

\begin{equation}
    \boldsymbol{U}^{n+1} = \frac{1}{3} \left(\boldsymbol{U}^{n}+2\boldsymbol{U}^{ **}+2\Delta \boldsymbol{U}^{**} \right),
\end{equation}
where at each stage, $\Delta \boldsymbol{U}$ is computed using the corresponding stage values.
For the magnetic field subroutine:

 \begin{equation}
     \boldsymbol{B}^{*}=\boldsymbol{B}^{n}+\Delta \boldsymbol{B}^n
 \end{equation}

\begin{equation}
    \boldsymbol{B}^{**}=\frac{1}{4} \left( 3\boldsymbol{B}^{n}+\boldsymbol{B}^{*}+\Delta \boldsymbol{B}^{*}\right)
\end{equation}

\begin{equation}
    \boldsymbol{B}^{n+1}=\frac{1}{3} \left(\boldsymbol{B}^{n}+2\boldsymbol{B}^{**}+2\Delta \boldsymbol{B}^{**}\right)
\end{equation}

In practice, a high-order time integrator, such as SSPRK3, allows a larger CFL number while still maintains stable and accurate solutions. Typically, $N_{\mathrm{CFL}} = 0.5$ is used for simulations with SSPRK3.

\subsection{Spatial Reconstruction}

To evaluate face fluxes, left and right interface states, $U^{L/R}$ and $B^{L/R}$, are computed by a reconstruction procedure. Low-order schemes are robust at discontinuities but diffusive, whereas high-order schemes are accurate for smooth flows but may generate oscillations near sharp gradients. Following Gamera, and LFM before it, we employ high-order spatial interpolation combined with the nonlinear Partial Donor Cell Method (PDM) limiter to preserve monotonicity. We retain this design philosophy while paying additional attention to consistency in curvilinear coordinates, as emphasized by \citet{monchmeyer1989,mignone2014}. Grid curvature must be accounted for; for example, the geometric and volumetric centers generally do not coincide. By incorporating curvature consistently, the enhanced PDM method (e-PDM) \citep{luo2025enhanced} attains very high-order convergence while maintaining monotonicity in arbitrary orthogonal curvilinear geometries. Here we specialize the formulation to the coordinate systems considered in this work, namely Cartesian, cylindrical, and spherical coordinates.

\subsubsection{The Reconstruction Step }

GAMERA-OP uses high-order upwind interpolation (fifth- or seventh-order) to obtain unlimited interface values from cell-averaged states. In Cartesian directions, reconstruction on uniform or non-uniform spacing follows the primitive-function approach \citep{Shu1998}. For a uniform Cartesian grid, the left-state coefficients are

 \begin{equation}
    \begin{aligned}
f_{i+\frac{1}{2}}^{L,5th}= & \frac{1}{30} f_{i-2}-\frac{13}{60} f_{i-1}
+\frac{47}{60} f_{i} \\ & +\frac{9}{20} f_{i+1}-\frac{1}{20} f_{i+2} ,
\end{aligned}
\end{equation}

\begin{equation}
    \begin{aligned}
f_{i+\frac{1}{2}}^{L,7th}= & -\frac{1}{140} f_{i-3}+\frac{5}{84} f_{i-2}-\frac{101}{420} f_{i-1}+\frac{319}{420} f_i \\
& +\frac{107}{210} f_{i+1}-\frac{19}{210} f_{i+2}+\frac{1}{105} f_{i+3} .
\end{aligned}
\end{equation}

Right-state values use the same coefficients with the stencil reversed.

In curvilinear directions ($R$ in cylindrical; $r,\theta$ in spherical), we adopt the conservative reconstruction of \citet{mignone2014}. We construct a degree-$(p-1)$ polynomial
\begin{equation}
f_i(\xi)=a_{i, 0}+a_{i, 1}\left(\xi-\xi_i^c\right)+a_{i, 2}\left(\xi-\xi_i^c\right)^2+\cdots+a_{i, p-1}\left(\xi-\xi_i^c\right)^{p-1}
\label{curvi polynomial}
\end{equation}
that satisfies the local conservation constraint
\begin{equation}
\langle{f}\rangle_i \ = \frac{1}{\Delta \mathcal{V}_i} \int_{\xi_{i-1/2}}^{\xi_{i+1/2}} f \, \partial_{\xi} \mathcal{V} \, d\xi.
\label{local conserv relation}
\end{equation}
with $\xi_i^c$ taken as zero for convenience. The one-dimensional Jacobian and the local cell volume in the sweep direction are

\begin{equation}
\partial_{\xi} \mathcal{V} = \begin{cases} R & \text { cylindrical, } \xi=R \\r^2 & \text { spherical, } \xi=r \\ \sin \theta & \text { spherical, } \xi=\theta\\ \ 1 &  \text {Cartesian-like direction} \ \xi \end{cases}
\end{equation}

\begin{equation}
\Delta \mathcal{V}_{\xi, i} = \begin{cases} \left(R_{i+\frac{1}{2}}^2-R_{i-\frac{1}{2}}^2\right) / 2 & \text { cylindrical, } \xi=R \\ \left( r_{i+\frac{1}{2}}^3-r_{i-\frac{1}{2}}^3 \right)/3 & \text { spherical, } \xi=r \\ \left( \cos \theta_{i-\frac{1}{2}}-\cos \theta_{i+\frac{1}{2}}\right) & \text { spherical, } \xi=\theta \\ \quad \quad \quad \Delta \xi &  \text {Cartesian-like direction} \ \xi\end{cases}
\end{equation}

Solving for the reconstruction polynomial leads to the following linear system: 
\begin{equation}
\begin{pmatrix}
\beta_{i-i_L,0} & \cdots & \beta_{i-i_L,p-1} \\
\vdots & \ddots & \vdots \\
\beta_{i+i_R,0} & \cdots & \beta_{i+i_R,p-1}
\end{pmatrix}^T
\begin{pmatrix}
w_{i,-i_L}^\pm \\
\vdots \\
w_{i,i_R}^\pm
\end{pmatrix}
=
\begin{pmatrix}
1 \\
\vdots \\
(\xi_{i\pm\frac{1}{2}} - \xi^c_i)^{p-1}
\end{pmatrix},
\end{equation}
where $i_L$ and $i_R$ denote the number of cells to the left and right of the central cell $i$ in the interpolation stencil, and $p = i_L + i_R + 1$ is the polynomial order.
The coefficients $\beta_{i+s,n}$ are geometry-dependent and given as follows:
\begin{equation}
\beta_{i+s,n} = \frac{1}{\Delta V_{i+s}} \int_{\xi_{i+s-\frac{1}{2}}}^{\xi_{i+s+\frac{1}{2}}} (\xi - \xi^c_i)^n \partial_{\xi} \mathcal{V}\, d\xi,
\end{equation}

and the interface states
\begin{equation}
f_{i+\frac{1}{2}}^L = \sum_{s=-i_L}^{i_R} w_{i,s}^+ \langle f \rangle_{i+s},
\end{equation}

\begin{equation}
f_{i-\frac{1}{2}}^R = \sum_{s=-i_L}^{i_R} w_{i,s}^- \langle f \rangle_{i+s}.
\end{equation}

\begin{figure}[htb!]
	\noindent\includegraphics[width= \textwidth]{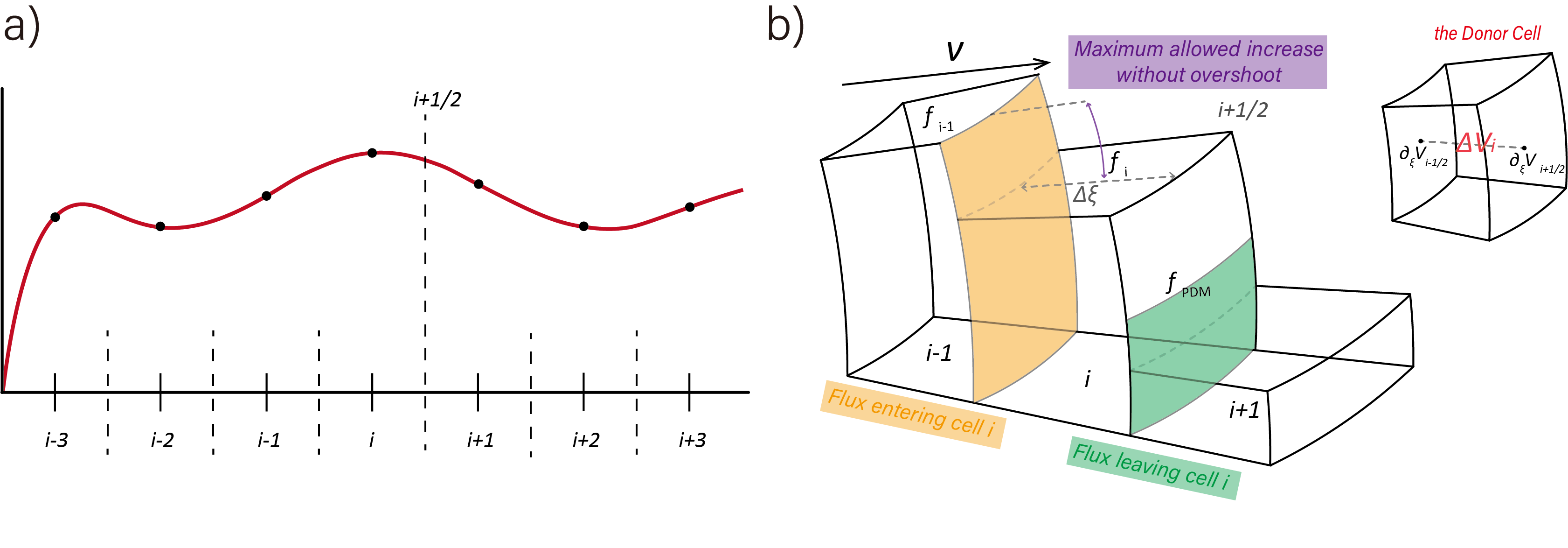}
	\centering
\caption{(a) A seventh-order reconstruction profile evaluating $f_{i+\frac{1}{2}}$. (b) A schematic showing the PDM limiter procedure for curvilinear coordinates.}\label{reconstruction_procedure}
\end{figure}

\begin{figure}[htb!]
	\noindent\includegraphics[width= 0.4 \textwidth]{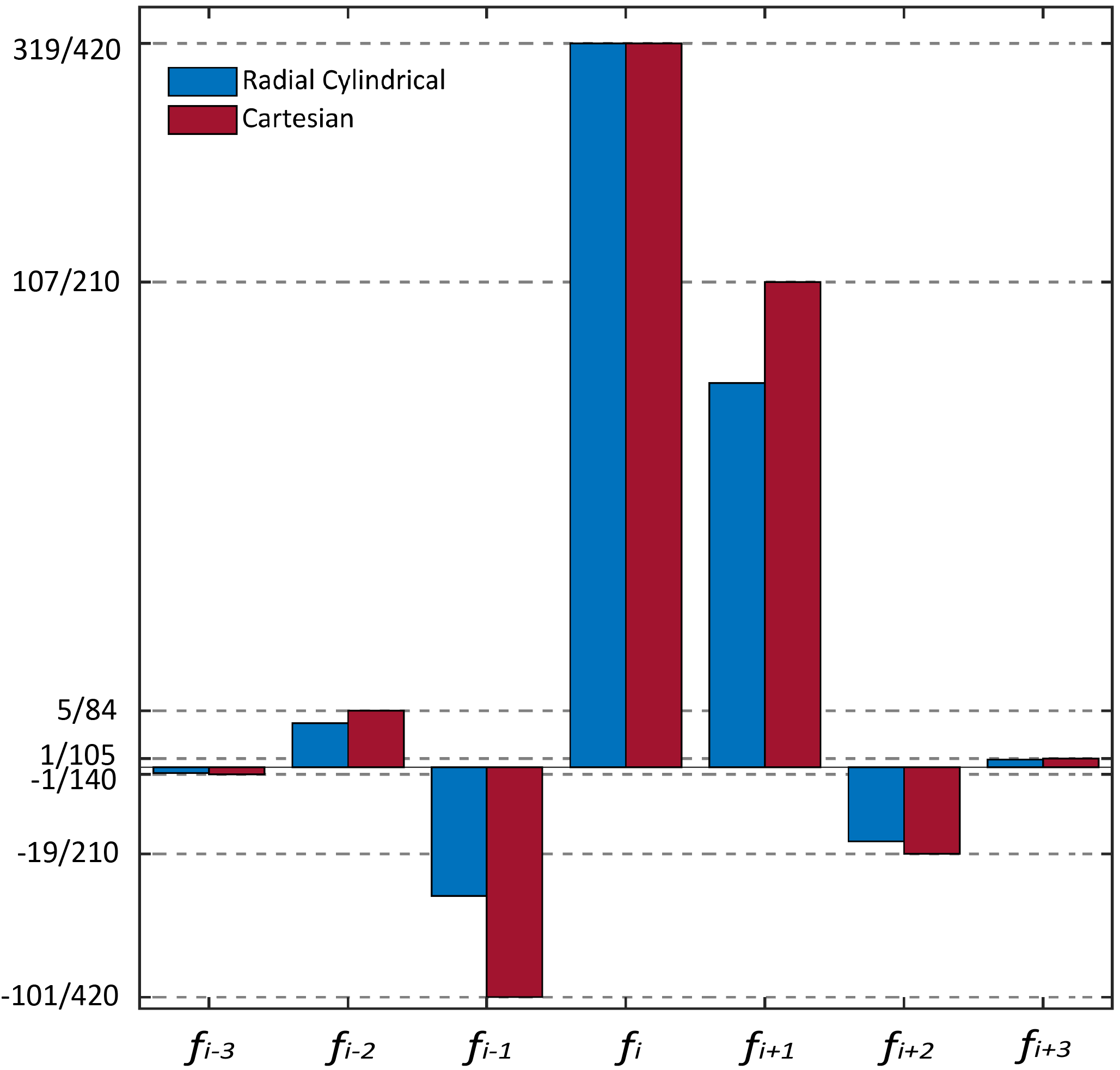}
	\centering
\caption{Seventh-order reconstruction weights for the second active face $f_{i+\frac{1}{2}}^L$ in a uniform cylindrical radial grid versus a Cartesian grid.}\label{weight compare plot}
\end{figure}

Figure \ref{reconstruction_procedure}a illustrates a seventh-order reconstruction of $f_{i+\frac{1}{2}}$. Using Cartesian weights in curved directions can be inaccurate \citep{monchmeyer1989,mignone2014}, as evident in the weight comparison shown in Figure \ref{weight compare plot} for a uniform cylindrical radial grid versus a Cartesian grid.

\subsubsection{The Limiting Step (e-PDM)}

After computing the unlimited interface state, we apply a limiter to prevent spurious overshoots and undershoots. We use the Partial Donor Cell Method (PDM),
which balances the in- and out-fluxes of a control volume to constrain the solution. Building on the original Cartesian formulation \citep{hain1987partial}, \citet{luo2025enhanced} incorporated the curvilinear volume element into the limiter. For the semi-discrete advection equation

\begin{equation}
    \frac{\partial f_i}{\partial t}=\frac{u_0}{\Delta \mathcal{V}_{\xi, i}}\left(\partial_{\xi} \mathcal{V}_{i-1 / 2} \cdot f_{i-1 / 2}-\partial_{\xi} \mathcal{V}_{i+1 / 2}\cdot f_{i+1 / 2}\right)
\end{equation}

the e-PDM limiter adjusts the high-order interface value only when necessary, ensuring that the update $f_i^n\to f_i^{n+1}$ does not create nonphysical extrema. The left state takes the form

\begin{equation}
    \begin{aligned}
f_{i+\frac{1}{2}}^L= & f_{i+\frac{1}{2}}^{\mathrm{Interp}*}-\operatorname{sign}\left(f_{i+1}-f_i\right) \cdot \max \\
& \left[0,  \vphantom{\Big|} \Big|f_{i+\frac{1}{2}}^{\mathrm{Interp}*}-f_i\Big|- C \cdot  \Big| \operatorname{sign}\left(f_i-f_{i-1}\right)\right. \\
& \left.+\operatorname{sign}\left(f_{i+1}-f_i\right)  \Big|\right],
\end{aligned}
\end{equation}

where $f_{i+\frac{1}{2}}^{\mathrm{Interp}*}$ is the high-order interpolated value confined between neighbors:
\begin{equation}
    f_{i+\frac{1} 
    {2}}^{\mathrm{Interp}*}=\operatorname{median}\left(f_i, f_{i+\frac{1}{2}}^{L,\mathrm{Interp}}, f_{i+1}\right),
\end{equation}
and $C$ is given by
\begin{equation}
C = \left(
\begin{aligned}
& a_1 \cdot\left| A \cdot \left( f_i - f_{i-1} \right) \right|+\\
& a_2 \cdot \left| 
\begin{aligned}
&   f_{i-1}\left( \dfrac{1}{1+2A} \partial_{\xi} \mathcal{V}_{i-\frac{1}{2}} - \dfrac{\Delta\mathcal{V}_{\xi, i}}{ \Delta \xi} \right) \bigg/ \left( \dfrac{2}{1+2A} \partial_{\xi} \mathcal{V}_{i+\frac{1}{2}} \right) \\
&\quad + f_{i}\left( \dfrac{\Delta\mathcal{V}_{\xi, i}}{ \Delta \xi} - \dfrac{1}{1+2A} \partial_{\xi} \mathcal{V}_{i+\frac{1}{2}} \right) \bigg/ \left( \dfrac{2}{1+2A} \partial_{\xi} \mathcal{V}_{i+\frac{1}{2}} \right)
\end{aligned}
\right|
\end{aligned}
\right)
\end{equation}

 where $A$ is the PDM parameter that controls the numerical diffusion and is related to the Courant number $N_{\mathrm{CFL}}$ \citep{hain1987partial,luo2025enhanced}. Setting $A=0$ the scheme recovers the diffusive first-order donor cell method, using $A>0$ yields the partial donor-cell limiter with reduced diffusion. We typically use $A=1$. The choice $(a_1,a_2)=(0,1)$ corresponds to the one-dimensional derivation, while optionally choosing values $(a_1,a_2)=(1/\left(A+1 \right),1-a_1)$ serves as additional dissipation for multi-dimensional correction.

Like other TVD-type limiters, the e-PDM also clips smooth extrema. We therefore provide an optional non-clipping switch that uses a five point stencil to distinguish true extrema from discontinuities \citep{leonard1991}. For the left state $f_{i+\frac{1}{2}}^L$, the unlimited value is retained if the curvature of a peak follows the shape of a local extremum:

$\begin{aligned}
    &1.  \quad \quad [(D_1 > 0) \& (D_2 > 0) \& (D_3 < 0) \& (D_4 < 0)] \\
    & \quad \mathrm{or} \quad [(D_1 < 0) \& (D_2 < 0) \& (D_3 > 0) \& (D_4 > 0)] \\
    &2. \quad \quad (|D_1| > |D_2|) \& (|D_3| < |D_4|)
\end{aligned}$\\
with
\begin{equation}
D_1 = \frac{f_{i-1} - f_{i-2}}{\langle x \rangle_{i-1}-\langle x \rangle_{i-2}}, \quad D_2 = \frac{f_i - f_{i-1}}{\langle x \rangle_{i}-\langle x \rangle_{i-1}}, \quad D_3 = \frac{f_{i+1} - f_i}{\langle x \rangle_{i+1}-\langle x \rangle_{i}}, \quad D_4 = \frac{f_{i+2} - f_{i+1}}{\langle x \rangle_{i+2}-\langle x \rangle_{i+1}}.
\end{equation}
Note the locations of volumetric centers $\langle x \rangle$ are considered when evaluating the local extrema indicator.

\subsection{Flux Functions}
\label{section flux}
Consider the following one-dimensional fluid equations, Eqs.~(\ref{continuity equaiton})--(\ref{entropy equaiton}), written with unity geometric scale factors:

\begin{equation}
\begin{aligned}
 \frac{\partial}{\partial t}\left[\begin{array}{c}
\rho \\
\rho u_{1} \\
\rho u_{2} \\
\rho u_{3} \\
\mathcal{E_P}\\
\mathcal{E_T}\\
S
\end{array}\right]=-\frac{\partial}{\partial x_1}& \left[\begin{array}{c}
\rho u_{1} \\
\rho u_{1}^2+P +B^2 / 2-B_{1}^2\\
\rho u_{1} u_{2} -B_{1} B_{2}\\
\rho u_{1} u_{3} -B_{1} B_{3}\\
u_{1}\left(\mathcal{E_P}+P\right)\\
u_1 \left(\mathcal{E_T}+{P} + {B^2}/{2}\right) -B_1\left( u_1 B_1+ u_2 B_2+u_3 B_3 \right) \\
S u_{1}
\end{array}\right] \\
& +\left[\begin{array}{c}
0 \\
0 \\
0 \\
u_{1} \frac{\partial}{\partial x_1}\left(B^2 / 2-B_{1}^2\right)+u_{2} \frac{\partial}{\partial x_2} B_{1} B_{2}+u_{3} \frac{\partial}{\partial x_3} B_{1} B_{3}\\
0\\
0
\end{array}\right],
\end{aligned}
\label{general 1d flux}
\end{equation}

The final term accounts for the work done by the Lorentz force, $\boldsymbol{u} \cdot \boldsymbol{F_L}$, in the plasma energy equation. Given reconstructed left and right interface states, the one-dimensional flux function provides the numerical fluxes across cell interfaces. In LFM and GAMERA, these numerical fluxes are evaluated using a Boltzmann-type solver (gas-kinetic scheme), which integrates the appropriate distribution function constructed from interface states in the velocity space. As high-order reconstruction is used, the choice of flux function has an insubstantial influence on solution accuracy ~\citep{lyon2004,mignone2010}. In practice, both gas-kinetic schemes and simpler alternatives, such as the Rusanov scheme~\citep{rusanov1961}, yield accurate results; therefore, we implement both options in GAMERA-OP.

\subsubsection{Gas Kinetic Scheme}

Unlike widely used approximate Riemann solvers, the gas-kinetic scheme computes numerical fluxes by integrating velocity-dependent distribution functions~\citep{xu1999}. In this approach, the fluid is represented as an ensemble of non-interacting particles that propagate ballistically over each time step. Macroscopic conserved quantities (mass, momentum, energy) are obtained by taking moments of the distribution, and these moments are used to reconstruct the updated distribution function for the next step.

The one-dimensional gas-kinetic fluxes of mass, momentum, and energy at the interface $(i+\frac{1}{2},j,k)$ are given by moment integrals of Maxwellian-based (magneto)gas-kinetic distribution functions constructed from the reconstructed left and right states, $(U^{L/R}, B^{L/R})$:

\begin{equation}
    f^{L/R}(v)=\sqrt{\frac{\rho^{L/R}}{2 \pi P^{L/R}}} e^{-\frac{\rho^{L/R}}{2 P^{L/R}}\left(v-u_{1}^{L/R}\right)^2}
\end{equation}

\begin{equation}
    f_B^{L/R}(v)=\sqrt{\frac{\rho^{L/R}}{2 \pi P_{\mathcal{T}}^{L/R}}} e^{-\frac{\rho^{L/R}}{2 P_{\mathcal{T}}^{L/R}}\left(v-u_{1}^{L/R}\right)^2}
\end{equation}

where $P_{\mathcal{T}} = P + B^2/2$ is the total pressure. To compute the fluxes, the positive (rightward) flux is obtained by integrating over $[0, \infty)$, and the negative (leftward) flux over $(-\infty, 0]$:

 \begin{equation}
     \boldsymbol{F}= \left(\begin{array}{c}
F_\rho \\
F_{\rho u_{1}} \\
F_{\rho u_{2}} \\
F_{\rho u_{3}} \\
F_{\mathcal{E_P}}\\
F_{\mathcal{E_T}}\\
F_{\mathcal{S}}
\end{array}\right)= \boldsymbol{F}^+ + \boldsymbol{F}^-,
 \end{equation}

 where
 
\begin{equation}
    \begin{aligned}
\boldsymbol{F}^+=&\left\langle v_B^1\right\rangle_{+}^L\left[\begin{array}{c}
\rho^L  \\
\rho^L u_{1}^L \\
\rho^L u_{2}^L \\
\rho^L u_{3}^L \\
0\\
\mathcal{E_T}^{{L}}+\frac{1}{2} (P_{\mathcal{T}}^L-B_1^2)\\
0
\end{array}\right]
+ \left\langle v_B^0\right\rangle_{+}^L\left[\begin{array}{c}
0 \\
P_{\mathcal{T}}^L-B_1^2 \\
-B_2^L B_1^L \\
-B_3^L B_1^L \\
0\\
\frac{1}{2} u_1^L (P_{\mathcal{T}}^L-B_1^2)-B_1 (u_2^L B_2^L+ u_3^L B_3^L) \\
0
\end{array}\right]\\
& +\left\langle v^1\right\rangle_{+}^L\left[\begin{array}{c}
0 \\
0 \\
0 \\
0\\
\mathcal{E_P}^L + \frac{1}{2} P^L\\
0\\
\frac{1}{2} P^L{(\rho^{L})^{1-\gamma }}
\end{array}\right]
+ \left\langle v^0\right\rangle_{+}^L\left[\begin{array}{c}
0 \\
0 \\
0 \\
0 \\
\frac{1}{2} u_1^L P^L\\
0 \\
\frac{1}{2} u_1^L P^L{(\rho^{L})^{1-\gamma }}
\end{array}\right] 
\end{aligned}
\end{equation}

The $B_1$ is already defined at the interface and does not need to be split into left and right states.

For smooth flows, i.e., $(\boldsymbol{U,B})^L=(\boldsymbol{U,B})^R$, the scheme recovers the true physical fluxes as other approximate Riemann solvers do. When the states are discontinuous across the interface, implicit diffusion is included in both the fluid and Lorentz stress numerical fluxes, differing from most approximate Riemann solvers, e.g., the Rusanov scheme.

\subsubsection{The Rusanov Flux}

GAMERA-OP also implements the Rusanov (local Lax-Friedrichs, LLF) scheme as an option for numerical fluxes. The Rusanov scheme consists of the arithmetic average physical flux and the diffusion term determined by the maximum wave speed. For the $x_1$ direction,
\begin{equation}
    F_1^{LLF} = \frac{F(U^L)+F(U^R)}{2} - \frac{a_{max}}{2} * ( U^R - U^L ) 
\end{equation}

where $F(U^{L/R})$ are the fluxes in (\ref{general 1d flux}) computed using the corresponding interface states, and $a_{max}$ is the local maximum signal propagation speed in $x_1$ direction estimated as:
\begin{equation}
    a_{}=\max \left(\left(\left|u_1^L\right|+c_f^L\right),\left(\left|u_1^R\right|+c_f^R\right)\right)
\end{equation}
with $c_f$ the fast mode speed.

\subsection{Constrained Transport}
\label{CT}

\subsubsection{Evolution of Magnetic Fields}

Applying Stokes’ theorem to Faraday’s law (\ref{faraday}) and integrating over each cell face yields the constrained-transport (CT) update for the face-centered magnetic-field components:

\begin{equation}
    \begin{aligned}
B_{1;i+\frac{1}{2}, j, k}^{ n+1}= & B_{1;i+\frac{1}{2}, j, k}^{n}- \frac{\Delta t}{A_{1;i+\frac{1}{2},j,k}}\left({E}_{2;i+\frac{1}{2}, j, k-\frac{1}{2}} \cdot L_{2;i+\frac{1}{2}, j, k-\frac{1}{2}}+{E}_{3;i+\frac{1}{2}, j+\frac{1}{2}, k} \cdot L_{3;i+\frac{1}{2}, j+\frac{1}{2}, k}\right. \\
& \left.-{E}_{2;i+\frac{1}{2}, j, k+\frac{1}{2}} \cdot L_{2;i+\frac{1}{2}, j, k+\frac{1}{2}}-{E}_{3;i+\frac{1}{2}, j-\frac{1}{2}, k} \cdot L_{3;i+\frac{1}{2}, j-\frac{1}{2}, k} \right),
\end{aligned}
\label{CT B1}
\end{equation}

\begin{equation}
    \begin{aligned}
B_{2;i, j+\frac{1}{2}, k}^{ n+1}= & B_{2;i, j+\frac{1}{2}, k}^{ n}-\frac{\Delta t}{A_{2;i,j+\frac{1}{2},k}}\left({E}_{3;i-\frac{1}{2}, j+\frac{1}{2}, k} \cdot L_{3;i-\frac{1}{2}, j+\frac{1}{2}, k}+{E}_{1;i, j+\frac{1}{2}, k+\frac{1}{2}} \cdot L_{1;i, j+\frac{1}{2}, k+\frac{1}{2}}\right. \\
& \left.-{E}_{3;i+\frac{1}{2}, j+\frac{1}{2}, k} \cdot L_{3;i+\frac{1}{2}, j+\frac{1}{2}, k}-{E}_{1;i, j+\frac{1}{2}, k-\frac{1}{2}} \cdot L_{1;i, j+\frac{1}{2}, k-\frac{1}{2}}\right) .
\end{aligned}
\label{CT B2}
\end{equation}

\begin{equation}
    \begin{aligned}
B_{3;i, j, k+\frac{1}{2}}^{n+1}= & B_{3;i, j, k+\frac{1}{2}}^{ n}-\frac{\Delta t}{A_{3;i,j,k+\frac{1}{2}}}\left({E}_{1;i, j-\frac{1}{2}, k+\frac{1}{2}} \cdot L_{1;i, j-\frac{1}{2}, k+\frac{1}{2}}+{E}_{2;i+\frac{1}{2}, j, k+\frac{1}{2}} \cdot L_{2;i+\frac{1}{2}, j, k+\frac{1}{2}}\right. \\
& \left.-{E}_{1;i, j+\frac{1}{2}, k+\frac{1}{2}} \cdot L_{1;i, j+\frac{1}{2}, k+\frac{1}{2}}-{E}_{2;i-\frac{1}{2}, j, k+\frac{1}{2}} \cdot L_{2;i-\frac{1}{2}, j, k+\frac{1}{2}}\right).
\end{aligned}
\label{CT B3}
\end{equation}

where $B_{1;i+\frac{1}{2}, j, k}$,$B_{2;i, j+\frac{1}{2}, k}$,$B_{3;i, j, k+\frac{1}{2}}$ denote the face-centered magnetic field components, and ${E}_{1;i, j\pm\frac{1}{2}, k\pm\frac{1}{2}}$,${E}_{2;i\pm\frac{1}{2}, j, k\pm\frac{1}{2}}$,${E}_{3;i\pm\frac{1}{2}, j\pm\frac{1}{2}, k}$ are the corresponding edge-centered electric fields.

These updates exactly conserve the magnetic flux through each cell. Moreover, if the initial condition satisfies $\nabla\cdot\boldsymbol{B}=0$, then the discrete divergence remains zero (to round-off) at all time steps:

\begin{equation}
\begin{aligned}
   \frac{\partial \left( \nabla \cdot \boldsymbol{B} \right)_{i,j,k}}{\partial t}=  \frac{1}{V_{i, j, k}} \Bigg[& \left(A_{1;i+\frac{1}{2}, j, k} \cdot \frac{\partial {B}_{1;i+\frac{1}{2}, j, k}}{\partial t} - A_{1;i-\frac{1}{2}, j, k} \cdot \frac{\partial {B}_{1;i-\frac{1}{2}, j, k}}{\partial t}\right) \\
& + \left(A_{2;i, j+\frac{1}{2}, k} \cdot \frac{\partial {B}_{2;i, j+\frac{1}{2}, k}}{\partial t} - A_{2;i, j-\frac{1}{2}, k} \cdot \frac{\partial {B}_{2;i, j-\frac{1}{2}, k}}{\partial t}\right) \\
& + \left(A_{3;i, j, k+\frac{1}{2}} \cdot \frac{\partial {B}_{3;i, j, k+\frac{1}{2}}}{\partial t} - A_{3;i, j, k-\frac{1}{2}} \cdot \frac{\partial {B}_{3;i, j, k-\frac{1}{2}}}{\partial t} \right) \Bigg]=0
\end{aligned}
\end{equation}

\subsubsection{Calculation of Electric Fields}

The construction of an edge-centered electric field is is based on values at edges. For example, to calculate of electric field on the edge located at $(i+\tfrac{1}{2},,j+\tfrac{1}{2},,k)$, we use values from the four neighboring control volumes $(i,j,k)$, $(i+1,j,k)$, $(i,j+1,k)$, and $(i+1,j+1,k)$.

At cell edges, fluid variable values, e.g., transverse velocities $u_1$, $u_2$, are obtained by two-step reconstruction (cell center $\rightarrow$ face $\rightarrow$ edge):
\begin{equation}
\boldsymbol{u}\left(i,j,k\right)\xrightarrow{\Delta x_1} \overline{\boldsymbol{u}} \left(i+\frac{1}{2},j,k\right)\xrightarrow{\Delta x_2}\overline{\boldsymbol{u}}\left(i+\frac{1}{2},j+\frac{1}{2},k\right),
\end{equation}

where the averaged values are defined as:
\begin{equation}
    \overline{\boldsymbol{u}} = \frac{1}{2} (\boldsymbol{u}^L+\boldsymbol{u}^R)
\end{equation}

The primary longitudinal magnetic fields are already located at cell interfaces and require a single reconstruction to the edge:
\begin{equation}
B_{1}\left(i,j+\frac{1}{2},k\right)\xrightarrow{\Delta x_2} \{ \overline{B}_{1},{B}_{1}^L,{B}_{1}^R \} \left(i+\frac{1}{2},j+\frac{1}{2},k\right).
\end{equation}

\begin{equation}
B_{2}\left(i,j+\frac{1}{2},k\right)\xrightarrow{\Delta x_1} \{\overline{B}_{2},{B}_{2}^L,{B}_{2}^R \} \left(i+\frac{1}{2},j+\frac{1}{2},k\right),
\end{equation}

with
\begin{equation}
    \overline{\boldsymbol{B}} = \frac{1}{2}(\boldsymbol{B}^L+\boldsymbol{B}^R)
\end{equation}

The electric field at cell edges is computed using a Rusanov-type scheme adapted to the edge geometry:

\begin{equation}
    \begin{aligned}
{E}_{3 ; i+ \frac{1}{2} , j + \frac{1}{2}, k }= & -\left( \overline{u\vphantom{B}}_1  \overline{B}_2 -\overline{u \vphantom{B}}_2  \overline{B}_1 \right) _{i+ \frac{1}{2} , j + \frac{1}{2}, k} \\
& +v_{D}\left(B_{2}^R-B_{2}^L+B_{1}^L-B_{1}^R\right) _{i+ \frac{1}{2} , j + \frac{1}{2}, k},
\end{aligned}
\label{E3 equation1}
\end{equation}

where $v_D$ is the local diffusion speed, given by:
\begin{equation}
    \begin{aligned}
v_{D;i+ \frac{1}{2} , j + \frac{1}{2}, k}=& \frac{1}{2}\left(|\boldsymbol{u}| + V_A\right)_{i+ \frac{1}{2} , j + \frac{1}{2}, k} \\
   =& \frac{1}{2} \left(\sqrt{ {u_{ 1}}^2+{u_{ 2}}^2}+ \sqrt{\left({B_{1}}^2+{B_{2}}^2\right)/ {\rho}}  \right) _{i+ \frac{1}{2} , j + \frac{1}{2}, k}.  
\end{aligned}
\end{equation}

The first term in Eq.~(\ref{E3 equation1}) approximates the smooth convective electric field, while the second term provides numerical diffusion, acting only across discontinuities. The magnitude of $v_D$ is typically set by the local flow and Alfv'en speed, but may be increased to the local fast mode speed for enhanced stability in specific problems.

\subsection{Curvilinear Coordinates}

In curvilinear coordinates, geometric source terms arise in the momentum equation from the divergence of the stress tensor. Accurate evaluation of these terms is essential for high-fidelity solutions. We present our finite‐volume approximations for cylindrical and spherical grids in Sections \ref{cylindrical source term} and \ref{spherical source term}, respectively.

Moreover, global simulations in cylindrical or spherical geometry contain an axis (or pole) singularity, posing two numerical challenges. First, degenerate faces and coincident cell edges at the axis lead to singular updates of the on‐axis magnetic field when using standard constrained‐transport. Second, the clustering of azimuthal cells near the axis imposes a prohibitively small CFL time step. In Section \ref{section axis} we describe the treatment of the addressed issues.

\subsubsection{Source Term Integration in Cylindrical Coordinate}
\label{cylindrical source term}

Expanding the momentum equation (\ref{momentum equation}) in cylindrical $(R,\phi,z)$ coordinates yields

\begin{equation}
   \frac{\partial \rho u_R}{\partial t}+\frac{1}{R} \frac{\partial\left(R M_{R R}\right)}{\partial R}+\frac{1}{R} \frac{\partial M_{\phi R}}{\partial \phi}+\frac{\partial M_{z R}}{\partial z}=\frac{M_{\phi \phi}}{R} ,
   \label{cylindrical radial momentum}
\end{equation}

\begin{equation}
    \frac{\partial \rho u_\phi}{\partial t}+\frac{1}{R} \frac{\partial\left(R M_{R \phi}\right)}{\partial R}+\frac{1}{R} \frac{\partial M_{\phi \phi}}{\partial \phi}+\frac{\partial M_{z \phi}}{\partial z}=-\frac{M_{R \phi}}{R},
    \label{cylindrical phi momentum}
\end{equation}

\begin{equation}
   \frac{\partial \rho u_z}{\partial t}+\frac{1}{R} \frac{\partial\left(R M_{R z}\right)}{\partial R}+\frac{1}{R} \frac{\partial M_{\phi z}}{\partial \phi}+\frac{\partial M_{z z}}{\partial z}=0,
\end{equation}

with $\overline{\boldsymbol{M}} $ the total stress tensor and $M_{i j}$ its component given as follows:

\begin{equation}
    \overline{\boldsymbol{M}} \equiv \rho \boldsymbol{u} \boldsymbol{u}-\boldsymbol{B} \boldsymbol{B}+\overline{\boldsymbol{I}} P_{*},
    \label{stress tensor general}
\end{equation}

\begin{equation}
    M_{i j} = \rho {u}_i {u}_j-{B}_i {B}_j + P_* \delta_{i j}.
    \label{stress tensor component}
\end{equation}

\textit{Radial Source Term}. The geometric source term $M_{\phi\phi}/R$ in (\ref{cylindrical radial momentum}) is evaluated by cell‐volume averaging

\begin{equation}
    \frac{M_{\phi \phi}}{R} = \frac{\int dR \int d\phi \int dz \left(\frac{M_{\phi \phi}}{R} R \right)} {\int dR \int d\phi \int dz R} =\left(\frac{A_{R; i+1 / 2}-A_{R; i-1 / 2}}{V} \right) M_{\phi \phi},
    \label{cylindnrical radial source term1}
\end{equation}

where $V=V_{i,j,k}$ and $M_{\phi \phi}=M_{\phi \phi;i,j,k}$.
This discretization preserves a constant-pressure static solution, as the source term exactly cancels the flux divergence.

\textit{Azimuthal Source Term. }This geometric source term in the $\phi$-momentum equation (\ref{cylindrical phi momentum}) is evaluated by linking the equation (\ref{cylindrical phi momentum}) to the angular momentum equation (\ref{cylindrical angular momentum}), which is more appropriate as no source term exists:

\begin{equation}
    \frac{\partial \rho u_\phi R}{\partial t}+\frac{1}{R} \frac{\partial\left(R^2 M_{R \phi}\right)}{\partial R}+\frac{1}{R} \frac{\partial\left( R M_{\phi \phi} \right)}{\partial \phi}+\frac{\partial \left( R M_{z \phi}\right)}{\partial z}=0
    \label{cylindrical angular momentum}
\end{equation}

The source terms are discretized as the difference between the two formulations (\ref{cylindrical phi momentum}) and (\ref{cylindrical angular momentum}):

\begin{equation}
    \frac{1}{R} M_{R \phi}=\frac{(R_{i+\frac{1}{2}}-R_{i-\frac{1}{2}})}{(R_{i+\frac{1}{2}}+R_{i-\frac{1}{2}}) V}\left(M_{R \phi; i+\frac{1}{2}} A_{R; i+\frac{1}{2}}+M_{R \phi; i-\frac{1}{2}} A_{R; i-\frac{1}{2}}\right) .
    \label{cylindrical phi source term}
\end{equation}

where $M_{R \phi; i\pm \frac{1}{2}}$ is the numerical flux at the corresponding face. This approach conserves angular momentum to round‐off, an especially important property in rotating systems, while preserving a uniform code structure with Cartesian implementations \citep{stone2020}.

\subsubsection{Source Term Integration in Spherical Coordinate}
\label{spherical source term}

In spherical $(r,\theta,\phi)$ coordinates, the expanded momentum equations read:

\begin{equation}
    \frac{\partial \rho u_r}{\partial t}  +\frac{1}{r^2} \frac{\partial\left(r^2 M_{r r}\right)}{\partial r} +\frac{1}{r \sin \theta} \frac{\partial(\sin \theta M_{\theta r})}{\partial \theta}+\frac{1}{r \sin \theta} \frac{\partial(M_{\phi r})}{\partial \phi} =\frac{1}{r} \left(M_{\theta \theta} +M_{\phi \phi} \right)
\end{equation}

\begin{equation}
    \begin{aligned}
\frac{\partial \rho u_\theta}{\partial t}+\frac{1}{r^2} & \frac{\partial\left(r^2 M_{r \theta}\right)}{\partial r}+\frac{1}{r \sin \theta} \frac{\partial \sin \theta M_{\theta \theta}}{\partial \theta}+\frac{1}{r \sin \theta} \frac{\partial M_{\phi \theta}}{\partial \phi} \\
& =- \frac{1}{r} \left(M_{\theta r} - \cot \theta M_{\phi \phi}\right)
\end{aligned}
\end{equation}

\begin{equation}
    \begin{aligned}
\frac{\partial \rho u_\phi}{\partial t}+\frac{1}{r^2} & \frac{\partial\left(r^2 M_{r \phi}\right)}{\partial r}+\frac{1}{r \sin \theta} \frac{\partial \sin \theta M_{\theta \phi}}{\partial \theta}+\frac{1}{r \sin \theta} \frac{\partial M_{\phi \phi}}{\partial \phi} \\
& =- \frac{1}{r}\left(M_{\phi r}+\cot \theta M_{\phi \theta}\right)
\label{spherical phi momentum}
\end{aligned}
\end{equation}

\textit{Radial Source Term}. Following (\ref{cylindnrical radial source term1}), the rhs of the $r$‐momentum equation is averaged by

\begin{equation}
    \frac{1}{r} \left(M_{\theta \theta} +M_{\phi \phi} \right) = \frac{A_{r; i+1 / 2}-A_{r; i-1 / 2}}{2V} \left(M_{\theta \theta} +M_{\phi \phi} \right).
\end{equation}

\textit{Meridional Source Term}. Analogous to (\ref{cylindnrical radial source term1}) and (\ref{cylindrical phi source term}), the source terms in the $\theta$- momentum equation are:

\begin{equation}
    \frac{1}{r} M_{r \theta}=\frac{(r_{i+\frac{1}{2}}-r_{i-\frac{1}{2}})}{(r_{i+\frac{1}{2}}+r_{i-\frac{1}{2}}) V}\left(M_{r \theta; i+\frac{1}{2}} A_{r; i+\frac{1}{2}}+M_{r \theta; i-\frac{1}{2}} A_{r; i-\frac{1}{2}}\right) ,
\end{equation}

\begin{equation}
    \frac{\cot \theta M_{\phi \phi}}{r}  = \left(\frac{A_{\theta; j+\frac{1}{2}}-A_{\theta; j-\frac{1}{2}}}{V} \right) M_{\phi \phi}
\end{equation}

\textit{Azimuthal Source Term}. As in cylindrical geometry, we subtract the $\phi$-momentum equation (\ref{spherical phi momentum}) from conservative $\phi$‐angular‐momentum form (\ref{spherical angular momentum}):

\begin{equation}
    \begin{aligned}
\frac{\partial \rho u_\phi r \sin \theta}{\partial t}+\frac{1}{r^2} & \frac{\partial\left(r^3 \sin \theta M_{r \phi}\right)}{\partial r}+\frac{1}{r \sin \theta} \frac{\partial \left (r \sin^2 \theta M_{\theta \phi} \right)}{\partial \theta}+\frac{1}{r \sin \theta} \frac{\partial \left( r \sin \theta M_{\phi \phi} \right) }{\partial \phi} \\
& =0
\label{spherical angular momentum}
\end{aligned}
\end{equation}

 The evaluation of the azimuthal source terms thus follows the same approach described in \citet{stone2020}. The expressions presented in \citet{stone2008} have typos by mistake; the corrected discrete forms are given as follows:

\begin{equation}
    \frac{1}{r} M_{r \phi}=\frac{(r_{i+\frac{1}{2}}-r_{i-\frac{1}{2}})}{(r_{i+\frac{1}{2}}+r_{i-\frac{1}{2}}) V}\left(M_{r \phi; i+\frac{1}{2}} A_{r; i+\frac{1}{2}}+M_{r \phi; i-\frac{1}{2}} A_{r; i-\frac{1}{2}}\right) .
\end{equation}

\begin{equation}
    \frac{\cot \theta M_{\phi \theta}}{r}=\frac{(\sin \theta_{j+\frac{1}{2}}-\sin \theta_{j-\frac{1}{2}})}{(\sin \theta_{j+\frac{1}{2}}+\sin \theta_{j-\frac{1}{2}}) V}\left(M_{\phi \theta; j+\frac{1}{2}} A_{\theta; j+\frac{1}{2}}+M_{\phi \theta; j-\frac{1}{2}} A_{\theta, j-\frac{1}{2}}\right) .
\end{equation}

\subsubsection{Polaraxis Issues}
\label{section axis}
 The computational domains of global simulations generally include the polar axis. The axis singularity occurs at $R=0$ in cylindrical or at $\theta=0,\pi$ in spherical coordinates (we do not consider the rarely used $r=0$ case). We focus on the cylindrical case - the spherical implementation is entirely analogous upon replacing $(R,\phi)$ with ($\theta,\phi)$. Boundary values in ghost cells across the axis are obtained by rotating active‐cell values by $180^\circ$ about the axis and applying a sign flip to vector components normal to the axis, that is, the $R$ and $\phi$ components on a cylindrical grid.
For example, a scalar $Q$ in cylindrical coordinates satisfies
\begin{equation}
    Q(1-n, j,k)=Q(n, \bmod (j+N_j / 2, N_j),k), n=1,2, \ldots N_g
\end{equation}

where $j=1\ldots N_j$ is the azimuthal index and $N_g$ is half the reconstruction stencil width.

Within constrained transport, on‐axis updates are modified for numerical stability. The axial electric field is set to the azimuthal average

\begin{equation}
    \mathcal{E}_z(i=1-\frac{1}{2}, j, k)=\frac{1}{N_\phi} \sum_{j=1}^{j=N_\phi} \mathcal{E}_z(i=1-\frac{1}{2}, j, k),
\end{equation}

and the on‐axis radial magnetic flux is not advanced directly by adjacent $\mathcal{E}$ values, instead it is reconstructed from neighbors,

\begin{align}
B_{R;1-\frac{1}{2}, j, k}^{ n+1}= & \frac{1}{2} \left( B_{R;1+\frac{1}{2}, j, k}^{n+1} + B_{R;1-\frac{3}{2}, j, k}^{n+1} \right).
\end{align}

\begin{figure}[htb!]
	\noindent\includegraphics[width=\textwidth]{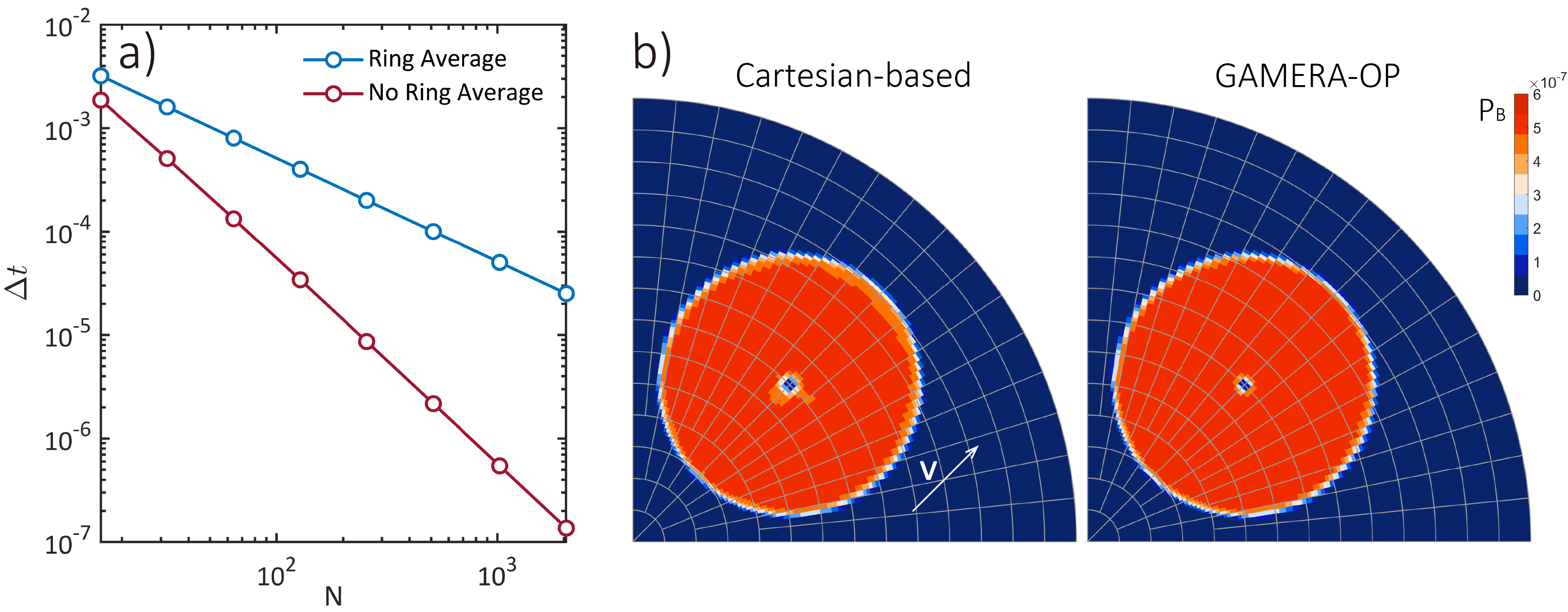}
	\centering
\caption{(a) Time step comparison for the two-dimensional field–loop advection problem on a cylindrical grid (see Section~\ref{loop 1/2D}).
(b) Magnetic energy distribution for the Cartesian–based method in the two-dimensional field–loop advection across the axis ($R=0$).
(c) Same as panel (b), computed with GAMERA-OP.}
\label{pole compare}
\end{figure}

To mitigate the severe CFL restriction caused by azimuthal cell clustering near the axis, we apply the Ring–Average technique \citep{zhang2019conservative}. In a small number of meridional “rings” closest to each pole, fluid variables are conservatively averaged over contiguous azimuthal chunks and then redistributed, which effectively increases the local azimuthal spacing. On the staggered magnetic grid, two face–normal components are averaged, and the third component is updated using the edge–centered $\mathcal{E}$ induced by the averaged flux changes, thereby preserving $\nabla\cdot\boldsymbol{B}=0$ both within each chunk and in every original cell. This procedure permits a substantially larger time step than the unmodified timestep, as shown in Figure~\ref{pole compare}a for the two dimensional field loop advection on a cylindrical grid. In addition, combined with the enhanced e-PDM reconstruction, GAMERA-OP markedly improves accuracy near the axis, as demonstrated in Figures~\ref{pole compare}b–c.

\subsection{Semi-relativistic Correction and Background Field Splitting}
Regions with strong background magnetic fields can have Alfv\'{e}n speeds that exceed the actual speed of light $c$. In an explicit scheme, this severely restricts the CFL time‐step and increases numerical diffusion. To alleviate these issues, we apply the semi‐relativistic (Boris) correction \citep{gombosi2002}. This method retains the displacement current term $\partial\boldsymbol{E}/\partial t$ in induction equation while using a reduced light speed $c'$. Full details appear in \citet{lyon2004,gombosi2002}. Effectively, the Alfvén speed is lowered to:

\begin{equation}
    V_{A}'= V_A/ \sqrt{1+V_A^2/{\hbox{$c'^2$}}}
\end{equation}

In many simulations with strong magnetic field gradient, it is also beneficial to split the magnetic field into $\boldsymbol{B}_{\mathrm{tot}}=\boldsymbol{B}+\boldsymbol{B}_0$, where $\boldsymbol{B}_0$ is an analytic background. The detailed implementation can be found in \citet{lyon2004,zhang2019}.

\subsection{Anisotropic MHD}

In collisionless or weakly collisional astrophysical plasmas, the assumption of scalar (isotropic) thermal pressure can be inaccurate, necessitating an anisotropic formulation. Accordingly, in the momentum equation (\ref{momentum equation}) the pressure tensor is generalized to the gyrotropic form:

\begin{equation}
    \overline{\boldsymbol{P}}=P_{\perp} \overline{\boldsymbol{I}}+\left(P_{\|}-P_{\perp}\right) \hat{\mathbf{b}} \hat{\mathbf{b}}
\end{equation}

where $\hat{\mathbf{b}}=\mathbf{B} /|\mathbf{B}|$ denotes the magnetic‐field unit vector, and $P_{\|}$and $P_{\perp}$ are are the pressures parallel and perpendicular to $\boldsymbol{B}$, respectively. The scalar pressure entering the energy equation is then defined by

\begin{equation}
    P=\frac{2 P_{\perp}+P_{\|}}{3} \label{average p}
\end{equation}

The evolution of the two pressure components follows from conservative polytropic relations \citep{hau2002}:

\begin{equation}
    \frac{\partial P_{\perp} B^{-\left(\gamma_{\perp} -1 \right)}}{\partial t}+\nabla \cdot\left(P_{\perp} B^{-\left(\gamma_{\perp} -1 \right)} \boldsymbol{u}\right)=0, \label{perp p}
\end{equation}

\begin{equation}
    \frac{\partial P_{\|}(B / \rho)^{\left(\gamma_{\|} -1 \right)}}{\partial t}+\nabla \cdot\left(P_{\|}(B / \rho)^{\left(\gamma_{\|} -1 \right)} \boldsymbol{u}\right)=0,  \label{para p}
\end{equation}

where $B=|\boldsymbol{B}|$ denotes the magnetic field strength and $\gamma_{\perp},\gamma_| $ are the polytropic indices. For the particular case  $\left( \gamma_{\perp},\gamma_| \right)= \left(2,3\right)$, the equations reduce to the well-known double adiabatic or Chew-Goldberger-Low (CGL) \citep{chew1956} system. 
 Together with the energy equation, relations (\ref{average p}), (\ref{perp p}), and (\ref{para p}) determine the total pressure and govern the partitioning of energy between the parallel and perpendicular components.

To enforce microphysical stability constraints, specifically the firehose, mirror, and ion‐cyclotron thresholds, a point‐implicit isotropization operator is applied whenever the anisotropy exceeds marginal limits. This relaxation term ensures that pressure anisotropy remains physically bounded. Numerical implementation details are provided in \citet{luo2023gas}.

\section{Tests}
\label{sec:tests}
\subsection{Linear Advection}

\begin{figure}[htb!]
	\noindent\includegraphics[width=\textwidth]{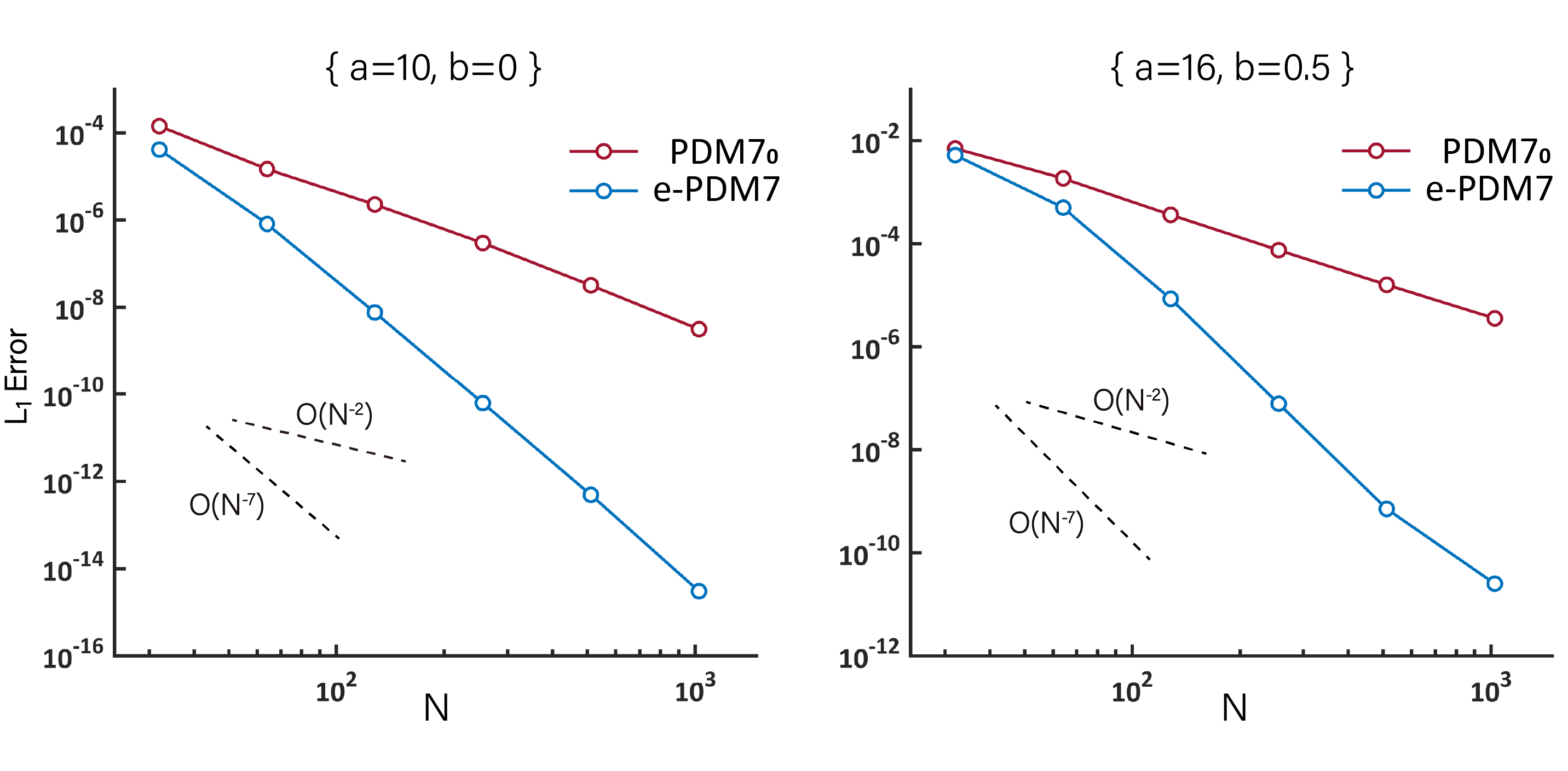}
	\centering
\caption{$L_1$ error of radial advection test as a function of grid resolution $N$}\label{1d_r}
\end{figure}

We adopt the scalar advection test proposed by \citet{mignone2014} as a one-dimensional linear benchmark to assess the effectiveness of our reconstruction and advection schemes. The initial condition is specified as a Gaussian profile, with parameters $a$ and $b$ determining the width and center of the Gaussian, respectively. Figure~\ref{1d_r} presents the convergence results for the cylindrical radial advection test, where the newly developed enhanced e-PDM method \citep{luo2025enhanced} is compared to the original seventh-order PDM ($\mathrm{PDM7}_0$) scheme. The enhanced e-PDM method consistently achieves its designed high-order accuracy, while the original seventh-order reconstruction ($\mathrm{PDM7}_0$) exhibits a reduction to second-order accuracy for this problem, as geometric effects are not properly accounted for in the original formulation.

\subsection{Rayleigh Rotational Stability Criterion}

Rayleigh’s criterion states that a rotating flow is stable to small, axisymmetric perturbations if $\partial_R \left[ (R^2 \Omega (R))^2\right]>0$. Experimental studies \citep{ji2006hydrodynamic} further indicate that flows satisfying Rayleigh’s criterion are stable not only to axisymmetric disturbances, but also to small, non-axisymmetric perturbations; such flows do not develop turbulence or facilitate significant angular momentum transport. For a differentially rotating flow of the form
\begin{equation}
    u_\phi(R)=R \Omega(R)=\Omega_0 R^{1-q},
    \label{differential rotation}
\end{equation}
Rayleigh’s criterion predicts that flows with $q<2$ remain stable, while those with $q>2$ violate the criterion and become unstable.

To assess the scheme's ability to conserve angular momentum and maintain rotational stability, we employ the rotational stability test of \citet{skinner2010}: the computational domain is $\left(R, \phi\right) \in [3, 7] \times [0, \pi/2]$, discretized with $200 \times 400$ cells. The fluid density and pressure are uniform, with $\rho=200$ and $P=1$ everywhere. A series of simulations is performed by introducing a uniformly distributed perturbation $\Delta \in [-10^{-4}, 10^{-4}]$ (see equation~\ref{perturbation uphi}) to the differentially rotating profile of equation~(\ref{differential rotation}), adopting $\Omega = 2\pi$ and varying $q$. A gravitational potential is added to maintain rotational equilibrium:
\begin{equation}
    \tilde{u}_\phi=u_\phi\left( 1+\Delta \right)
    \label{perturbation uphi}
\end{equation}

Simulations are performed for several values of $q$, specifically $q \in {1,1.5,1.95,1.99,2.01,2.05}$, with $q=1$ corresponding to galactic disk rotation and $q=1.5$ corresponding to Keplerian rotation. Each simulation is run to $t = 300$ (i.e., 300 orbits). To quantify the development of instability, we use the scaled stress:
\begin{equation}
    \frac{\langle R \rho u_R \delta u_\phi \rangle}{\langle RP \rangle} = \frac{\iint R \rho u_R (u_\phi - R \Omega) R dR d\phi}{\iint R P R dR d\phi}.
    \label{scaled stress}
\end{equation}

\begin{figure}[htb!]
\noindent\includegraphics[width=\textwidth]{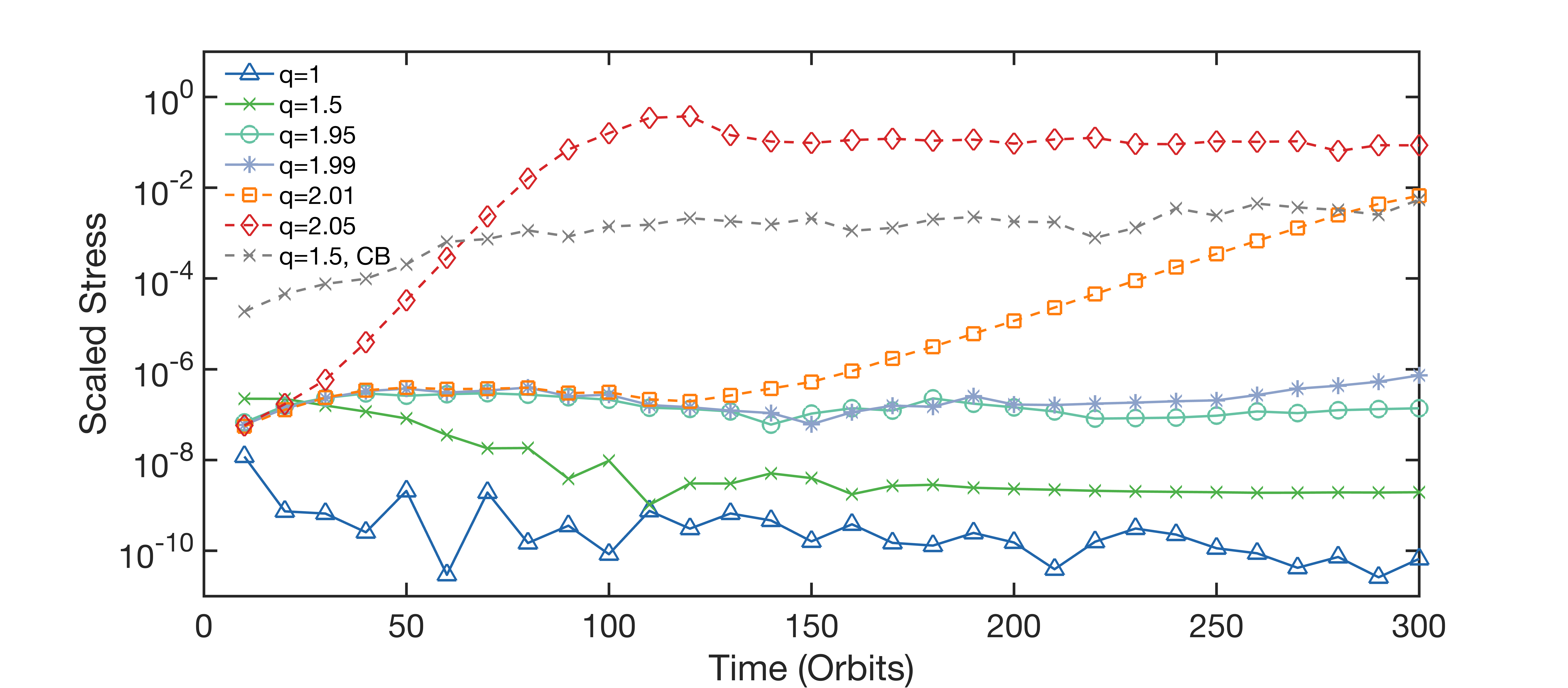}
\centering
\caption{Rotational stability test. Time evolution of the scaled stress for several values of $q$.}\label{RL2D-2}
\end{figure}

Figure~\ref{RL2D-2} shows the time evolution of the scaled stress for different $q$ values. As expected, flows with $q<2$ remain stable, while those with $q>2$ rapidly become unstable. To further illustrate the importance of angular momentum conservation, we repeat the $q=1.5$ case using a Cartesian-based algorithm as in \citet{zhang2019}, mapping the problem onto a cylindrical grid. For this method, truncation error in angular momentum leads to rapid instability within a few orbits, causing the simulation to crash as negative pressure develops. To maintain positive pressure throughout, the initial value of $P$ is increased to $50$, corresponding to a temperature seven times higher. The result, shown by the grey line in Figure~\ref{RL2D-2}, demonstrates that the instability quickly saturates (for $t<5$), leading to rapid mass loss from the domain.

\subsection{Spherical Wind}

We present the spherical wind hydrodynamics test case from \citet{mignone2014}, formulated in two-dimensional cylindrical coordinates ($R$–$z$), to assess the effectiveness of the numerical schemes in solving multidimensional nonlinear problems in curvilinear geometry.  The computational domain is defined as $\left(R, z\right) \in [0, 10] \times [-10, 10]$, discretized on a grid of $N_R \times N_z = 256 \times 512$ cells. The initial condition consists of a steady, spherically symmetric wind/blast wave embedded within a static background medium, prescribed as

\begin{align}
    \left(\rho, u_R,u_z,P\right)= \begin{cases}(\frac{1}{u_r r^2}, \tanh(5 r) \frac{R}{r},\tanh(5 r) \frac{z}{r},\frac{c_{s,w}^2 \rho^\gamma}{\gamma}) &  r = \sqrt{R^2 + z^2} \leq 1 \\ (\frac{1}{4}, 0,0,\frac{c_{s,a}^2 \rho}{\gamma}) & r = \sqrt{R^2 + z^2} > 1 \end{cases}
\end{align}

where $c_{s,w} = 3 \times 10^{-2}$ and $c_{s,a} = 4 \times 10^{-3}$ denote the wind and ambient sound speeds, respectively. Within the region $r \leq 1$, the physical variables remain fixed, ensuring that the supersonic wind propagates outward continuously. At $R = 0$, a ‘pole’ boundary condition is enforced, while symmetry conditions are imposed on all other boundaries.

\begin{figure}[htb!]
	\noindent\includegraphics[width=\textwidth]{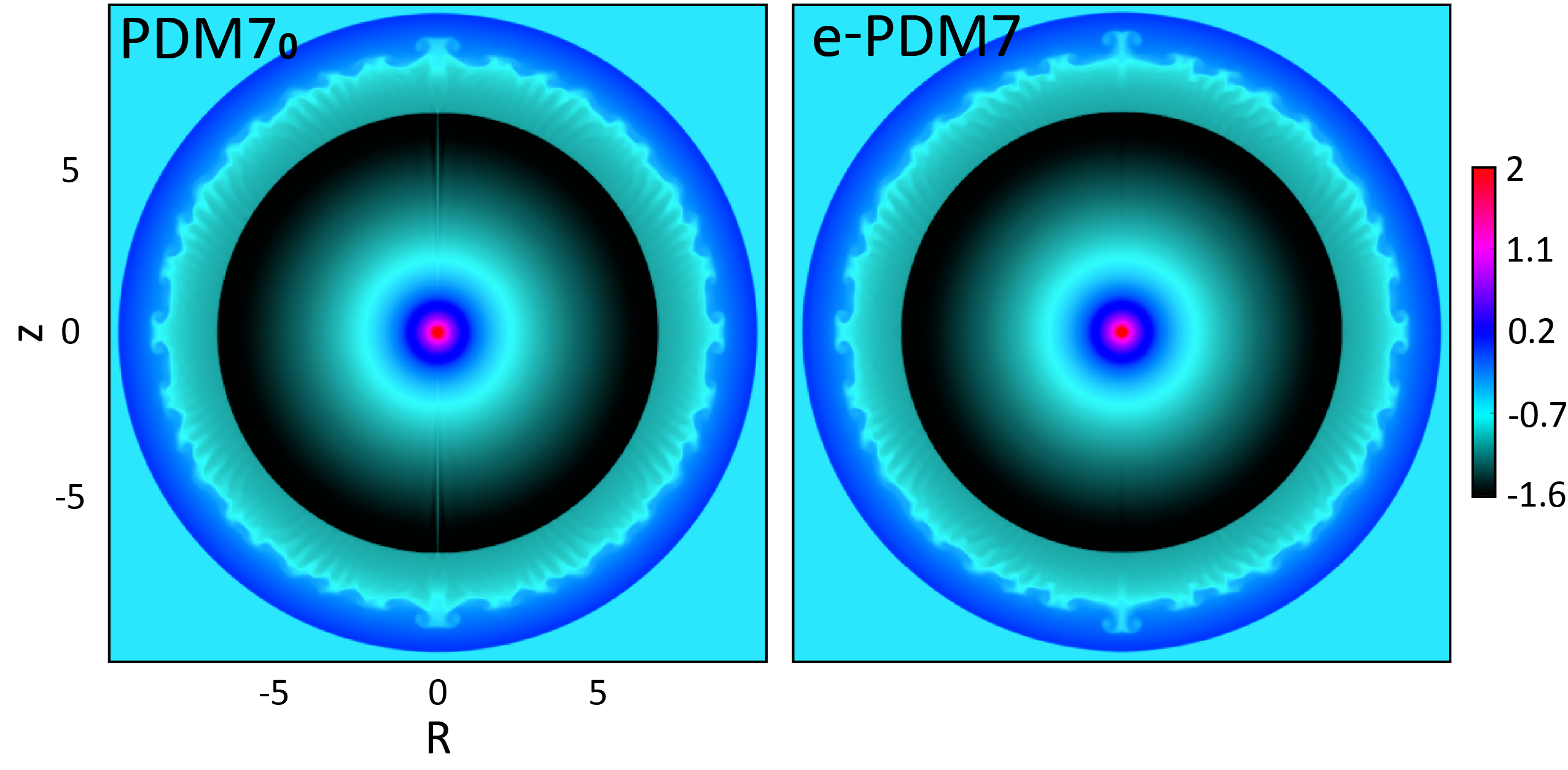}
	\centering
\caption{The spatial density of log density at $t=20$ in the spherical wind test, comparing results from the uncorrected $\mathrm{PDM7}_0$ (left) and enhanced $\mathrm{PDM7}$ (right) schemes.}\label{WIND2D}
\end{figure}

As shown by \citet{mignone2014}, numerical schemes that do not properly account for geometric curvature effects can generate unphysical overshoots along the symmetry axis. Figure~\ref{WIND2D} compares the solution obtained using the original PDM7 method without geometry correction ($\mathrm{PDM7}_0$, left panel) to that obtained by the enhanced e-PDM7 method (right panel). The improved scheme not only accurately resolves the spherically symmetric flow but also captures fine-scale structures near the contact wave arising from Rayleigh–Taylor instabilities, highlighting its low numerical diffusion and high-resolving power.

\subsection{Orszag-Tang Vortex}

We perform the standard Orszag–Tang vortex test \citep{orszag1979small} to evaluate the scheme’s ability to resolve complex MHD shock interactions and turbulence. The initial conditions are set with uniform density and pressure, given by $\rho = 25/(36\pi)$ and $p = 5/(12\pi)$, respectively. The velocity field is initialized as $u_{x} = -\sin(2\pi y)$ and $u_{y} = \sin(2\pi x)$, and the magnetic field is specified through the vector potential $A_{z} = B_{0} \left[ \cos(4\pi x)/(4\pi) + \cos(2\pi y)/(2\pi) \right]$ with $B_{0} = 1$. The computational domain is $\left(x, y\right) \in [0, 1]^2$, and periodic boundary conditions are imposed in both directions.

\begin{figure}[htb!]
\noindent\includegraphics[width=\textwidth]{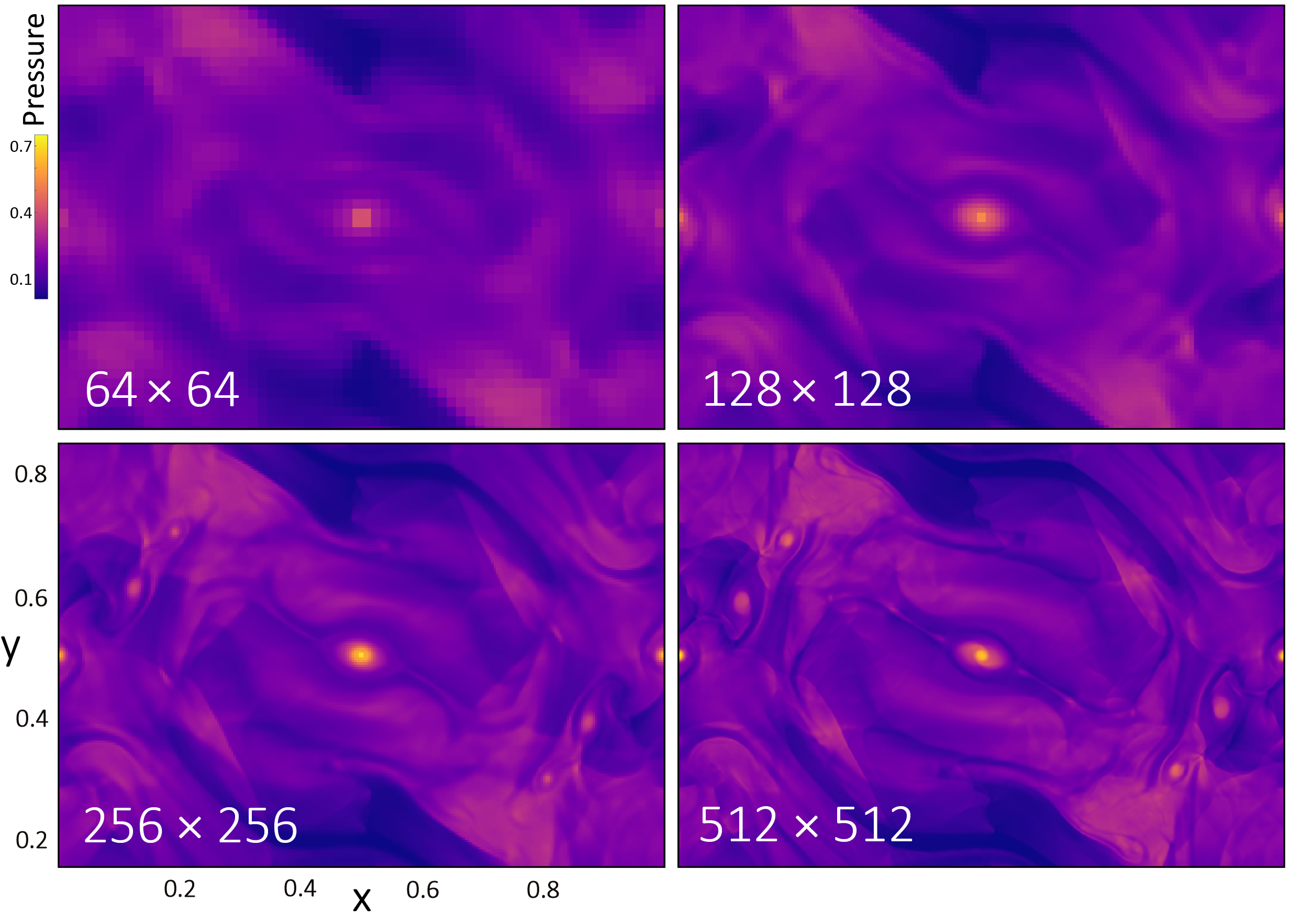}
\centering
\caption{Spatial distribution of plasma pressure $P$ at $t = 1$ in the Orszag–Tang vortex problem for four different grid resolutions.}\label{OT2D_2}
\end{figure}

At early times, the solution remains smooth. As the system evolves, the formation and interaction of supersonic shocks lead to the development of complex turbulent structures. The appearance of fine-scale features, such as the central bead/vortex, at late times is an indicator of the scheme’s ability to minimize numerical diffusion. Figure~\ref{OT2D_2} displays the spatial distribution of plasma pressure at $t = 1$ for a sequence of increasingly refined grids, namely $64^2$, $128^2$, $256^2$, and $512^2$. The solutions exhibit no spurious oscillations, which demonstrates the robustness of the algorithm for handling strong shock interactions. The central vortex is well resolved even at the coarsest resolution of $64^2$, which highlights the accuracy and resolving power of the MHD solver.

\subsection{MHD Rotor}

We carry out the MHD rotor test \citep{balsara1999staggered,mignone2007}. This problem consists of a dense, uniformly rotating fluid disk at the center, embedded in a stationary, low-density ambient medium. The transition between the disk and ambient fluid is smoothed using a linear taper. An initially uniform magnetic field is imposed perpendicular to the rotation axis. The resulting velocity shear at the fluid interface produces strong rotational discontinuities.

The initial conditions are given by

\begin{align}
    \left(\rho, u_R,u_\phi,P\right)= \begin{cases}(10, 0,\omega R,1) & \text { for } R<R_0 \\ \left(1+9 f, 0,f \omega R_0,1\right) & \text { for } R_0 \leq R \leq R_1 \\ (1,0,0,1) & \text { otherwise }\end{cases}
\end{align}

where $\omega = 20$, $R_0 = 0.1$, and the taper function $f = (R_1 - R)/(R_1 - R_0)$ connects the rotating core at $R_0$ to the ambient fluid at $R_1 = 0.115$. The ratio of specific heats is set to $\gamma = 1.4$, and the initial magnetic field is uniform, $B_x = B_0 = 5/\sqrt{4\pi}$.

\begin{figure}[htb!]
\noindent\includegraphics[width=\textwidth]{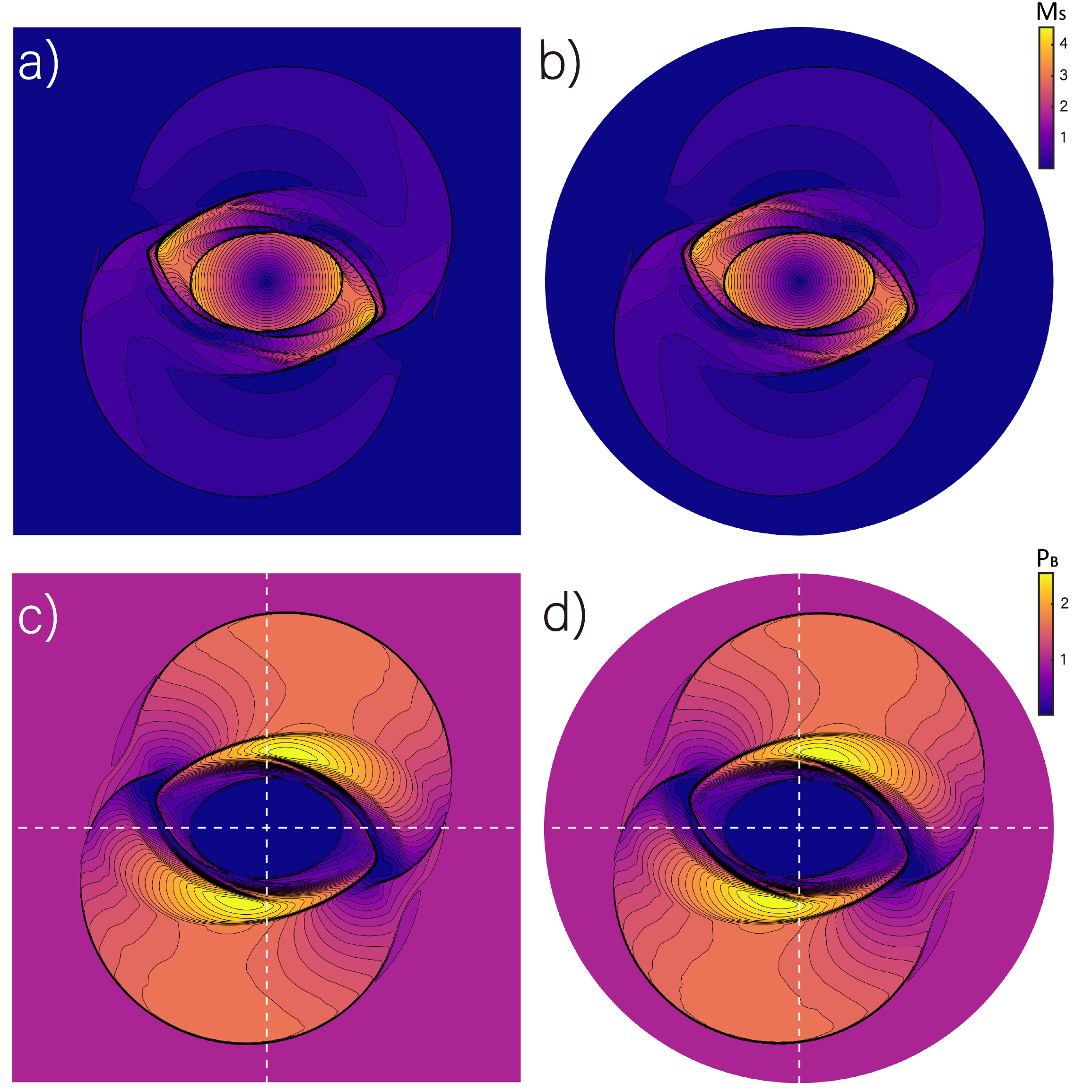}
\centering
\caption{Spatial distribution of Mach number (top) and magnetic pressure (bottom) at $t = 0.15$ in MHD rotor simulations using Cartesian coordinates (left, $512 \times 512$ cells) and cylindrical coordinates (right, $256 \times 512$ cells).}\label{ROTOR2D_1}
\end{figure}

\begin{figure}[htb!]
\noindent\includegraphics[width=\textwidth]{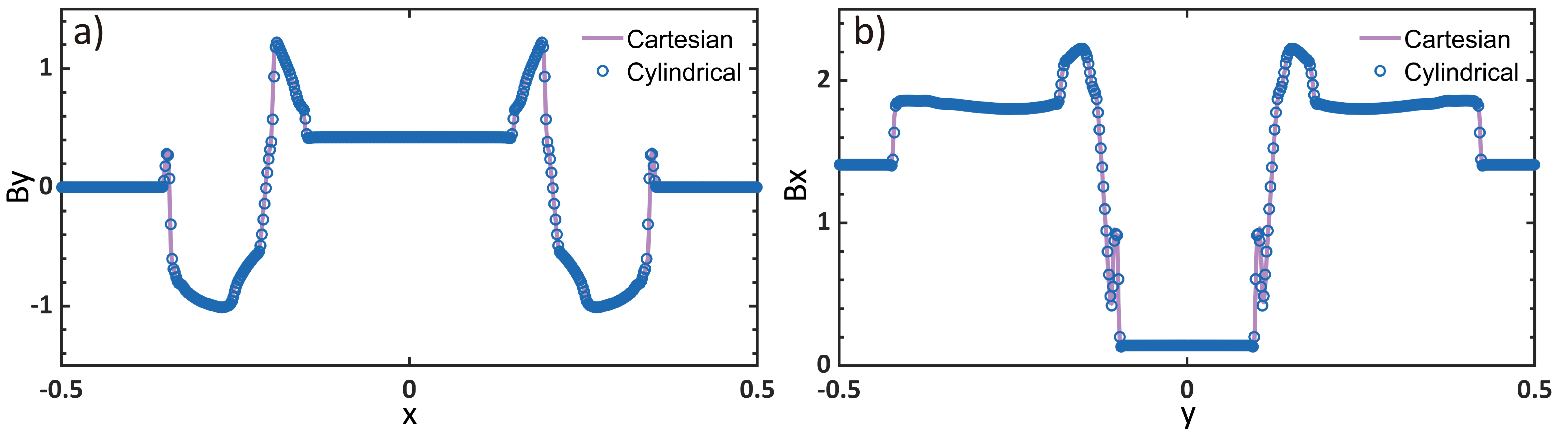}
\centering
\caption{Line profiles of $B_y$ along $y = 0$ and $B_x$ along $x = 0$ at $t = 0.15$ in MHD rotor simulations.}\label{ROTOR2D_2}
\end{figure}

Simulations are performed using both Cartesian and cylindrical coordinate systems. For the Cartesian case, the grid covers $\left(x, y\right) \in [-0.5, 0.5]^2$ with a resolution of $512 \times 512$ cells. In cylindrical coordinates, the grid spans $\left(R, \phi\right) \in [0, 0.5] \times [0, 2\pi]$ at a resolution of $256 \times 512$. Figure~\ref{ROTOR2D_1} presents the spatial distribution of the Mach number and magnetic pressure at $t = 0.15$ for both geometries. The agreement between Cartesian and cylindrical results confirms the consistency of the numerical method. The central region exhibits well-resolved concentric contours of Mach number. Figure~\ref{ROTOR2D_2} displays line profiles (taken along white dashed lines in Figure~\ref{ROTOR2D_1}) of the magnetic field components, $B_y$ along $x = 0$ and $B_x$ along $y = 0$. The profiles are nearly indistinguishable between the two coordinate systems, demonstrating the high degree of symmetry preservation and the absence of numerical oscillations.

\subsection{ Field Loop Advection}
\subsubsection{1D/2D Advection of a Field Loop}
\label{loop 1/2D}
We employ the magnetic field loop advection test originally proposed by \citet{gardiner2005unsplit} in both Cartesian and cylindrical geometries. This test serves as an MHD analog of multi-dimensional square-wave advection and provides a stringent assessment of the accuracy of the MHD solver's constrained transport (Maxwell) subroutine. In this setup, the plasma density, velocity, and pressure are initially uniform, with $\rho = 1$ and $P = 1$. The magnetic field is initialized as a weak field loop defined by the vector potential
\begin{equation}
    A_z =  \operatorname{max}\left( \left[ B_0 \left( a_0-r \right)\right] \right)
\end{equation}


where $B_0 = 10^{-3}$ specifies the magnetic field strength, $a_0 = 0.3$ is the loop radius, and $r$ denotes the distance from the loop center. Since the plasma is high-$\beta$, the magnetic field is expected to advect passively with the flow.

\begin{figure}[htb!]
\noindent\includegraphics[width=\textwidth]{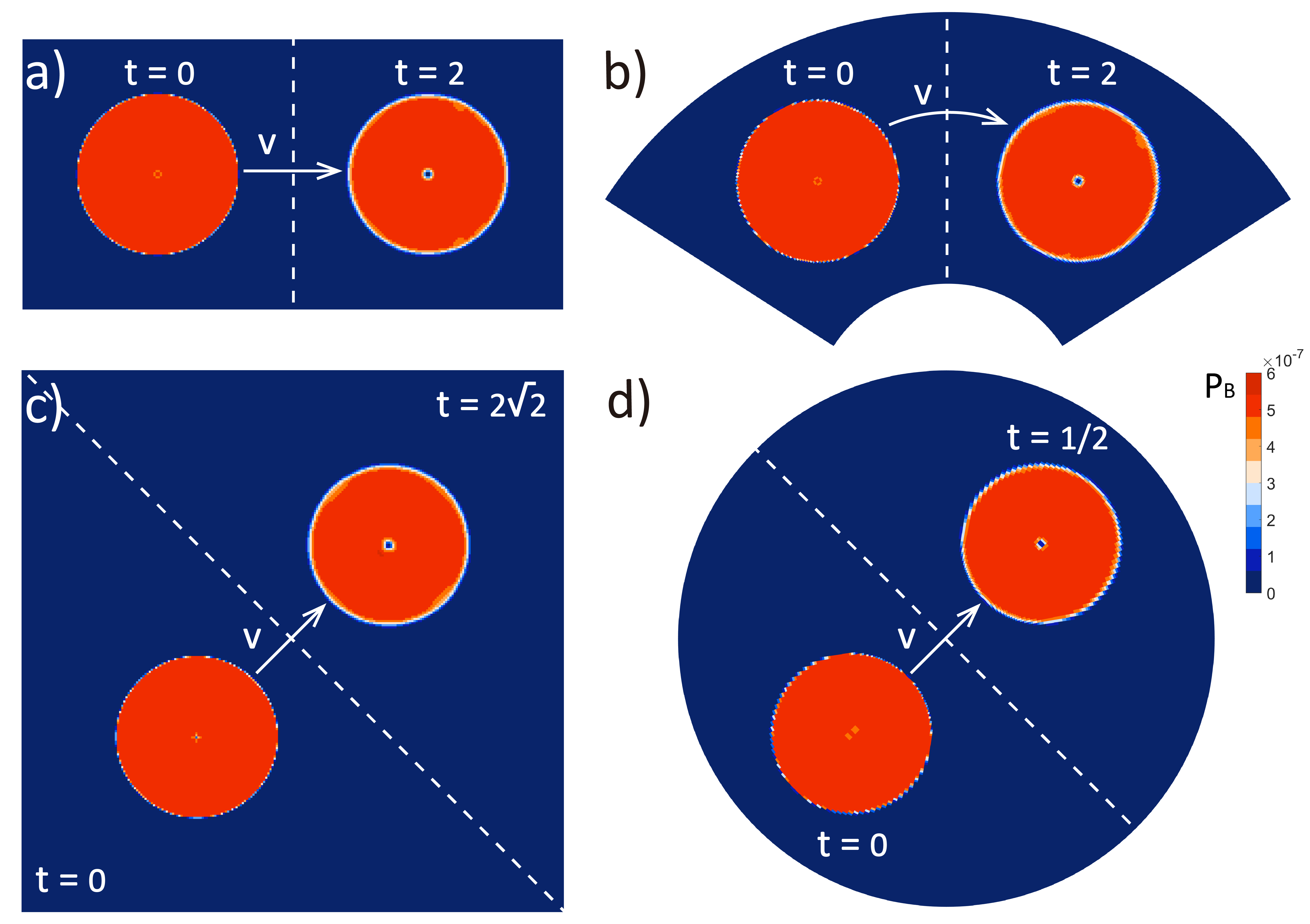}
\centering
\caption{Spatial distribution of magnetic energy from four field-loop advection simulations. Panels (a) and (b) show the results of one-dimensional advection using Cartesian and cylindrical coordinates, respectively. Panels (c) and (d) display two-dimensional advection in Cartesian and cylindrical coordinates. Field loop positions have been offset for visual clarity.}\label{LOOP2D_1}
\end{figure}

We first consider advection by a one-dimensional velocity field. In Cartesian coordinates, the domain is $\left(x, y\right) \in [-1, 1] \times [-0.5, 0.5]$ with $256 \times 128$ cells and velocity $(u_x, u_y) = (1, 0)$. The corresponding cylindrical setup uses $\left(R, \phi\right) \in [0.5, 1.5] \times [\pi/2 - 1, \pi/2 + 1]$, with a $128 \times 256$ grid and azimuthal velocity $u_\phi = \Omega_0 R$ with $\Omega_0 = 1$. Periodic boundary conditions are applied in both cases. A similar configuration is described in \citet{skinner2010}. Figure~\ref{LOOP2D_1}(a)-(b) displays the magnetic energy after the loop completes two transits of the domain ($t = 2$) compared to the initial state. The nearly perfect preservation of the loop shape and sharp edges, without spurious oscillations, demonstrates the high fidelity of the constrained transport algorithm.
 
We next perform the test with a two-dimensional velocity field. In Cartesian coordinates, the grid covers $\left(x, y\right) \in [-1, 1]^2$ with $256^2$ cells, and the velocity is set to $(u_x, u_y) = (\sqrt{2}, \sqrt{2})$ so that the loop advects twice across the domain by $t = 2\sqrt{2}$. In cylindrical geometry, the grid spans $\left(R, \phi\right) \in [0, 1] \times [0, 2\pi]$ at $128 \times 256$ resolution. The velocity field is defined as $u_R = \sqrt{2}[\cos(\phi) + \sin(\phi)]$, $u_\phi = \sqrt{2}[\cos(\phi) - \sin(\phi)]$. The magnetic field loop is initially centered at $(R_0, \phi_0) = (0.5, 5\pi/4)$, so the loop advects diagonally across the domain, traversing the $R = 0$ singularity at $t = 0.5$. Figure~\ref{LOOP2D_1}(c)-(d) show the results for the two-dimensional test. The field loop retains its integrity even as it passes through the pole, confirming the robustness of the reconstruction, constrained transport, and boundary conditions.

\begin{figure}[htb!]
\noindent\includegraphics[width=\textwidth]{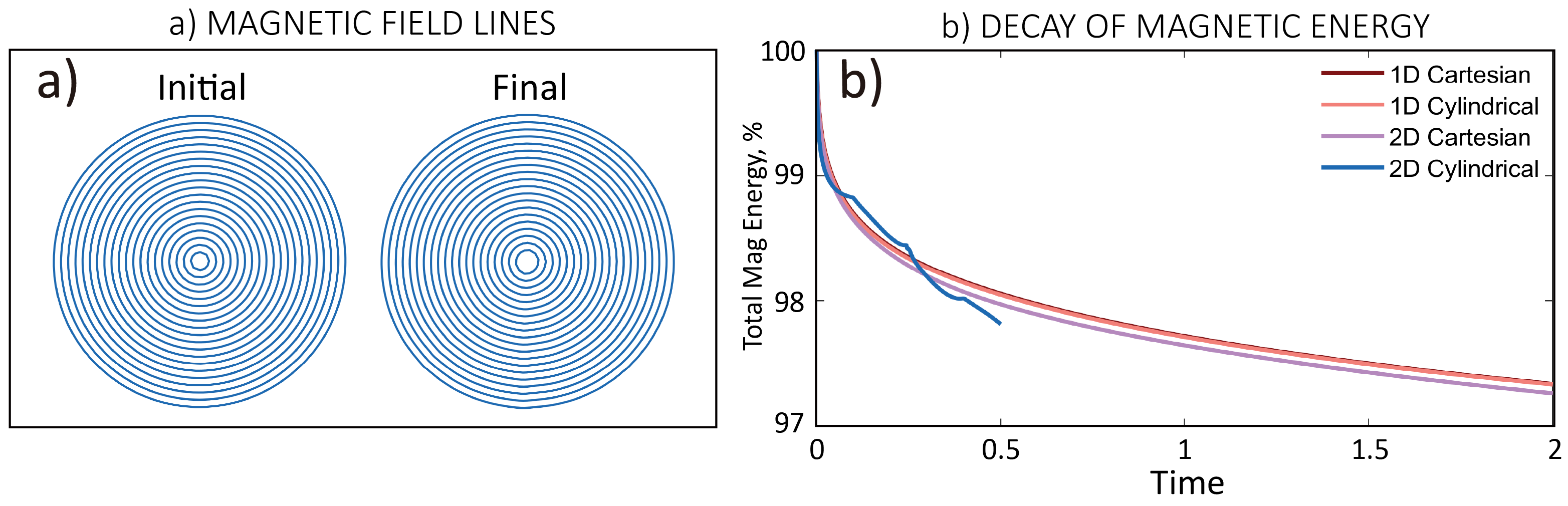}
\centering
\caption{(a) Magnetic field lines at $t = 0$ and $t = 0.5$ in the two-dimensional cylindrical advection problem. (b) Time evolution of total magnetic energy in all four simulations.}\label{LOOP2D_2}
\end{figure}

Figure~\ref{LOOP2D_2}(a) shows magnetic field lines at $t = 0$ and $t = 0.5$ in the cylindrical two-dimensional advection problem. The loop remains nearly perfectly circular, without visible distortion, even in this challenging scenario. To quantify numerical diffusion, Figure~\ref{LOOP2D_2}(b) presents the volume-integrated magnetic energy as a function of time. For the one-dimensional advection, the Cartesian and cylindrical results are indistinguishable, with over $98 \%$ of the initial energy retained after two periods. The two-dimensional Cartesian case exhibits only slightly higher diffusion. In the cylindrical case with advection across the pole, the decay of magnetic energy does not strictly follow a power-law due to grid anisotropy along the advection path, but overall diffusion remains low and there is no evidence of significant overshoot or undershoot.

\subsubsection{3D Advection of a Field Loop}

We employ a three-dimensional version of the field loop advection test, following \citet{stone2020}, to evaluate the accuracy and robustness of the numerical algorithm in a fully three-dimensional setting. A magnetic loop with finite thickness, centered at $x_0 = -\sqrt{2}/2$, $y_0 = 0$, and $z_0 = \sqrt{2}/2$, is initialized via the vector potential
\begin{equation}
\begin{aligned}
A_z = B_0 \exp \left[-\frac{(z-z_0)^2}{\sigma^2}\right] \max\left(R - \sqrt{(x-x_0)^2 + (y-y_0)^2},, 0\right),
\end{aligned}
\end{equation}
where $B_0 = 10^{-3}$ is the magnetic field strength, $R = 0.5$ is the loop radius, and $\sigma = 0.2$ is the thickness parameter. The velocity, density, and pressure are initialized uniformly as $u_x = 1$, $\rho = 1$, and $p = 0.5$. The computational domain is a half-sphere defined by $\left(R,\theta,\phi \right) \in [0.1, 2] \times [0, \pi/2] \times [0, 2\pi]$, discretized with $192 \times 128 \times 256$ cells in $r$, $\theta$, and $\phi$, respectively.

\begin{figure}[htb!]
	\noindent\includegraphics[width=\textwidth]{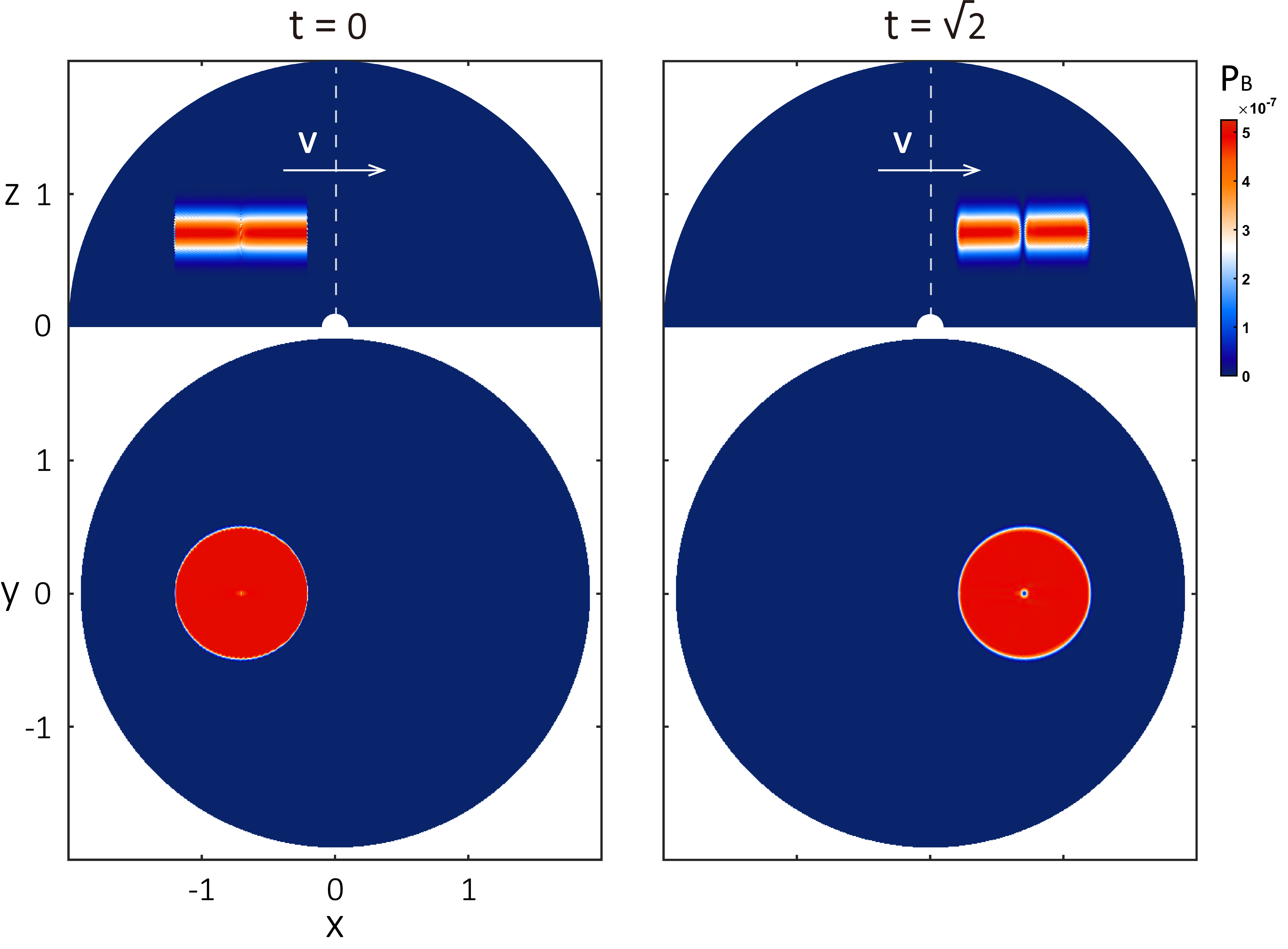}
	\centering
\caption{Spatial distribution of magnetic pressure in the three-dimensional magnetic field loop advection test in spherical coordinates. Results are shown for the $z = \sqrt{2}/2$ plane (top) and $y = 0$ plane (bottom) at $t = 0$ (left) and $t = \sqrt{2}$ (right).}\label{LOOP3D}
\end{figure}

This test presents a stringent challenge for the scheme, as the magnetic loop is advected across the coordinate singularity at $\theta = 0$. Figure~\ref{LOOP3D} shows the spatial distribution of magnetic pressure on the $z = \sqrt{2}/2$ and $y = 0$ planes at $t = \sqrt{2}$. The results exhibit minimal distortion or diffusion, demonstrating the effectiveness of the numerical method in three-dimensional curvilinear geometry, even in the presence of a coordinate singularity.

\subsection{MHD Blast Wave}

 \subsubsection{2D Blast Wave}
We perform two-dimensional MHD blast wave simulations to assess the robustness of the numerical schemes in capturing strong, multidimensional MHD shocks and discontinuities.

The test is conducted in both Cartesian and cylindrical geometries. The initial conditions consist of a uniform density, zero velocity, and a high-pressure region within a central circle, while the pressure is low elsewhere:
\begin{equation}
    \left(\rho, u_{R}, u_{\phi}, P,\right)= \begin{cases}(1,0,0,0.1) & (R> 0.1) \\ (1,0,0,10) & (R\leq 0.1)\end{cases}
\end{equation}
The initial magnetic field is oriented along the $x$-direction, with $B_x = B_0 = 1$. In cylindrical geometry, a divergence-free magnetic field is initialized using the vector potential $A_z = B_0 R \sin \phi$.

\begin{figure}[htb!]
\noindent\includegraphics[width=\textwidth]{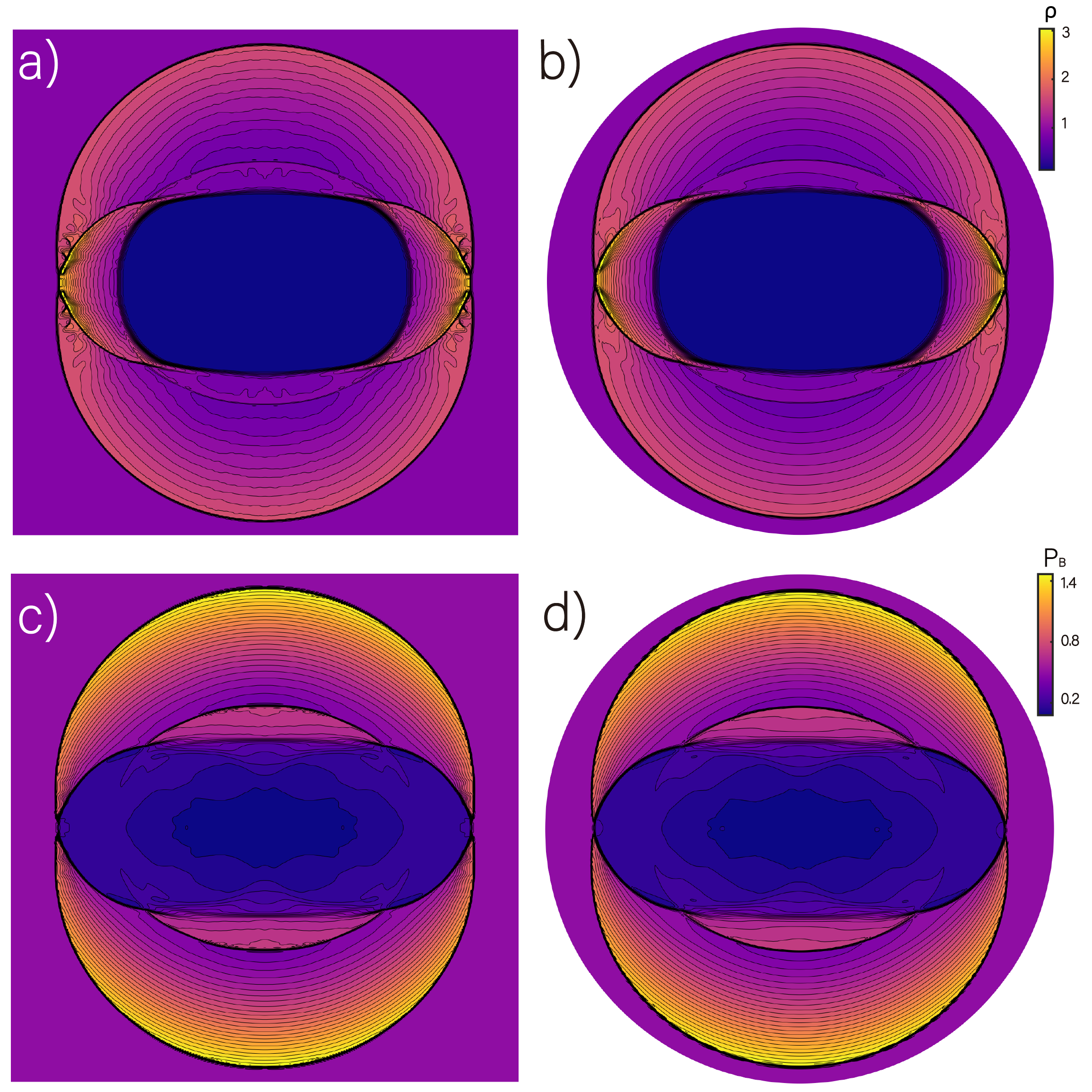}
\centering
\caption{Spatial distribution of plasma density (top) and magnetic pressure (bottom) at $t = 0.18$ in two-dimensional MHD blast wave simulations. Results are shown for Cartesian coordinates (left, $512 \times 512$ cells) and cylindrical coordinates (right, $256 \times 512$ cells).}\label{BW2D_1}
\end{figure}

Figure~\ref{BW2D_1} presents the spatial distributions of plasma density and magnetic pressure at $t = 0.18$, using a computational domain of $(x, y) \in [-0.5, 0.5]^2$ with $512^2$ grid cells for the Cartesian case and $(R, \phi) \in [0, 0.5] \times [0, 2\pi]$ with a $256 \times 512$ grid for the cylindrical case. The results in both geometries are in close agreement, with the shock structures, rarefactions, and symmetry accurately resolved.

\subsubsection{3D Blast Wave}
A three-dimensional version of the blast wave test is carried out in both Cartesian and spherical geometries. The Cartesian simulation uses a domain $(x, y, z) \in [-0.5, 0.5]^3$ with $192^3$ cells. The spherical simulation adopts $(R, \theta, \phi) \in [0.5, 1.5] \times [\pi/2 - \pi/5, \pi/2 + \pi/5] \times [-\pi/5, \pi/5]$, also with $192^3$ cells. In spherical coordinates, the divergence-free magnetic field aligned with the $x$-direction is initialized via the vector potentials $A_r=B_0/\sqrt{2} r \sin\theta \cos \theta  (\sin \phi -\cos \phi)$, $A_\theta=-B_0/\sqrt{2} r \sin^2 \theta (\sin \phi -\cos \phi)$ in spherical coordinate.

\begin{figure}[htb!]
\noindent\includegraphics[width=\textwidth]{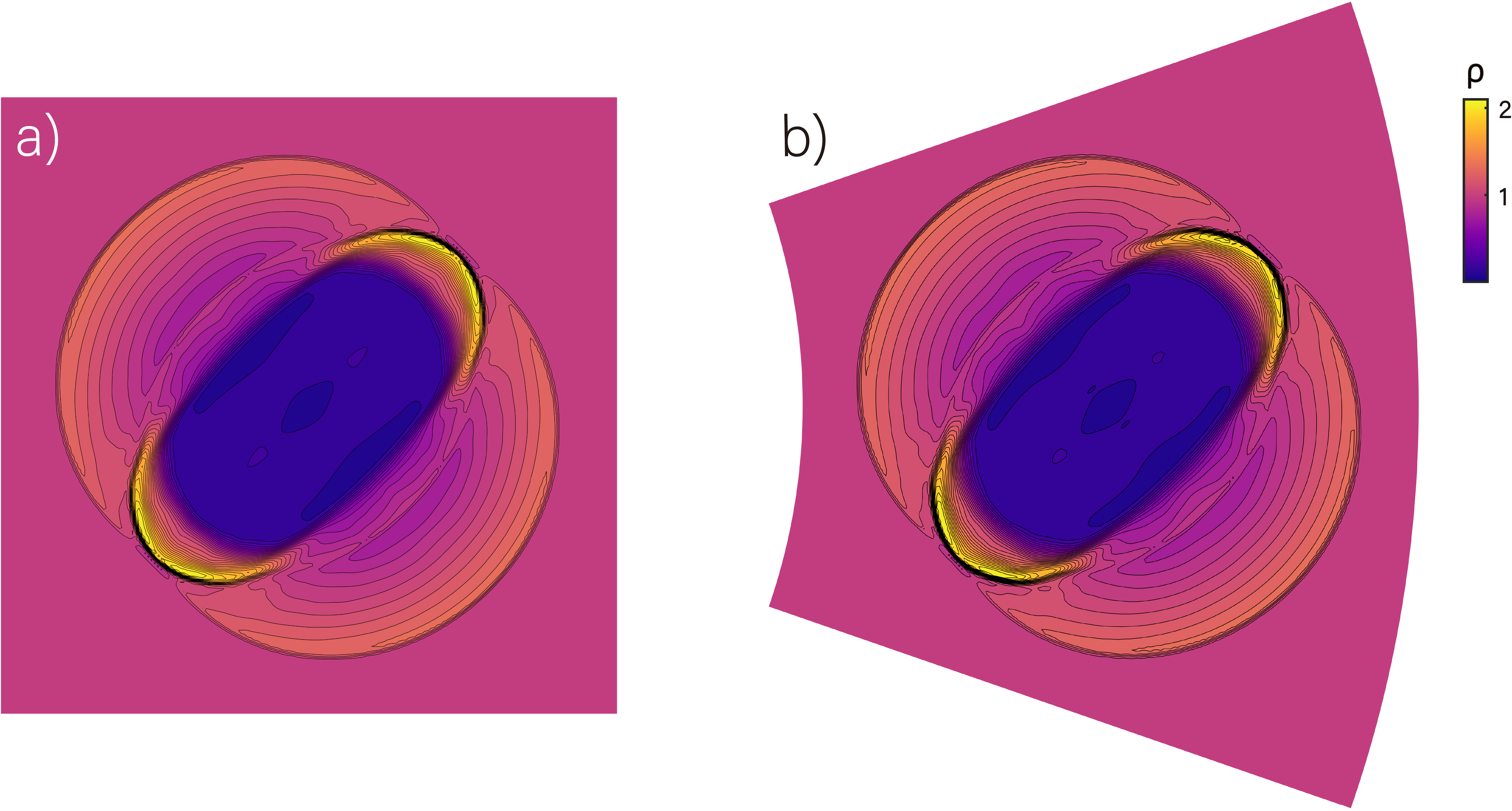}
\centering
\caption{Spatial distribution of plasma density at $t = 0.18$ in the $x$-$y$ plane at $z = 0$ (top) for three-dimensional MHD blast wave simulations using Cartesian coordinates (left, $192^3$ cells) and spherical coordinates (right, $192^3$ cells).}\label{BW3D_2}
\end{figure}

Figure~\ref{BW3D_2} displays the spatial distribution of plasma density at $t = 0.18$ in the $x$-$y$ plane at $z=0$ for both geometries. The solutions demonstrate excellent agreement, and in particular, the spherical grid maintains symmetry well despite the grid anisotropy and spatially varying background field.

\section{Summary}
\label{sec:summary}

We have introduced GAMERA-OP (Orthogonal-Plus), a new code that is reinvention of LFM/GAMERA designed to solve the ideal (and extended) MHD system on arbitrary orthogonal curvilinear geometry. The code advances face-centered magnetic fields with constrained transport, employs geometry-consistent high-order reconstruction with an enhanced PDM limiter, and offers flexible numerical flux and time integrator option. In cylindrical and spherical coordinates, we use angular-momentum-preserving source term discretizations and a ring-average procedure to relax CFL constraints while preserving $\nabla\cdot\mathbf{B}=0$. For stiff Alfvénic regimes, the code includes the semi-relativistic (Boris) correction and background-field splitting. The code also supports anisotropic (gyrotropic) pressure closures for various space and astrophysical plasma problems.

A broad test suite in Cartesian, cylindrical, and spherical geometries demonstrates the code's accuracy and robustness. Test results shows solution of minimal diffusion and numerical oscillation, regardless of the computational geometry employed. For many stringent tests, GAMERA-OP demonstrates superior performance to the original GAMERA algorithm.

GAMERA-OP is an evolving platform: beyond the schemes presented here, new physics modules, performance optimizations, and numerical upgrades are being integrated on an ongoing basis. This paper serves as a reference for users and developers who wish to understand the current implementation or tailor it to specific problems. The code’s modular architecture facilitates straightforward setup and interchangeable modules, supporting fast iteration and easy extensibility.

\acknowledgments

We thank Christopher White and James Stone for helpful discussions.
This research was supported by the NSFC General Program (42374216) and the RGC General Research Fund (RFS2526-7S05, 17309224 and 17309725).

\bibliography{GAMERA_OP}{}

@ARTICLE{Gombosi2002,
       author = {{Gombosi}, Tamas I. and {T{\'o}th}, G{\'a}bor and {De Zeeuw}, Darren L. and {Hansen}, Kenneth C. and {Kabin}, Konstantin and {Powell}, Kenneth G.},
        title = "{Semirelativistic Magnetohydrodynamics and Physics-Based Convergence Acceleration}",
      journal = {Journal of Computational Physics},
         year = 2002,
        month = mar,
       volume = {177},
       number = {1},
        pages = {176-205},
          doi = {10.1006/jcph.2002.7009},
       adsurl = {https://ui.adsabs.harvard.edu/abs/2002JCoPh.177..176G}
}

@ARTICLE{stone2008,
       author = {{Stone}, James M. and {Gardiner}, Thomas A. and {Teuben}, Peter and {Hawley}, John F. and {Simon}, Jacob B.},
        title = "{Athena: A New Code for Astrophysical MHD}",
      journal = {\apjs},
         year = 2008,
        month = sep,
       volume = {178},
       number = {1},
        pages = {137-177},
          doi = {10.1086/588755},
archivePrefix = {arXiv},
       eprint = {0804.0402},
 primaryClass = {astro-ph},
       adsurl = {https://ui.adsabs.harvard.edu/abs/2008ApJS..178..137S}
}

@ARTICLE{skinner2010,
       author = {{Skinner}, M. Aaron and {Ostriker}, Eve C.},
        title = "{The Athena Astrophysical Magnetohydrodynamics Code in Cylindrical Geometry}",
      journal = {\apjs},
         year = 2010,
        month = may,
       volume = {188},
       number = {1},
        pages = {290-311},
          doi = {10.1088/0067-0049/188/1/290},
archivePrefix = {arXiv},
       eprint = {1004.2487},
 primaryClass = {astro-ph.IM},
       adsurl = {https://ui.adsabs.harvard.edu/abs/2010ApJS..188..290S}
}

@ARTICLE{stone2020,
       author = {{Stone}, James M. and {Tomida}, Kengo and {White}, Christopher J. and {Felker}, Kyle G.},
        title = "{The Athena++ Adaptive Mesh Refinement Framework: Design and Magnetohydrodynamic Solvers}",
      journal = {\apjs},
         year = 2020,
        month = jul,
       volume = {249},
       number = {1},
          eid = {4},
        pages = {4},
          doi = {10.3847/1538-4365/ab929b},
archivePrefix = {arXiv},
       eprint = {2005.06651},
 primaryClass = {astro-ph.IM},
       adsurl = {https://ui.adsabs.harvard.edu/abs/2020ApJS..249....4S}
}

@ARTICLE{mignone2007,
       author = {{Mignone}, A. and {Bodo}, G. and {Massaglia}, S. and {Matsakos}, T. and {Tesileanu}, O. and {Zanni}, C. and {Ferrari}, A.},
        title = "{PLUTO: A Numerical Code for Computational Astrophysics}",
      journal = {\apjs},
         year = 2007,
        month = may,
       volume = {170},
       number = {1},
        pages = {228-242},
          doi = {10.1086/513316},
archivePrefix = {arXiv},
       eprint = {astro-ph/0701854},
 primaryClass = {astro-ph},
       adsurl = {https://ui.adsabs.harvard.edu/abs/2007ApJS..170..228M}
}

@ARTICLE{zhang2019,
       author = {{Zhang}, Binzheng and {Sorathia}, Kareem A. and {Lyon}, John G. and {Merkin}, Viacheslav G. and {Garretson}, Jeffrey S. and {Wiltberger}, Michael},
        title = "{GAMERA: A Three-dimensional Finite-volume MHD Solver for Non-orthogonal Curvilinear Geometries}",
      journal = {\apjs},
         year = 2019,
        month = sep,
       volume = {244},
       number = {1},
          eid = {20},
        pages = {20},
          doi = {10.3847/1538-4365/ab3a4c},
archivePrefix = {arXiv},
       eprint = {1810.10861},
 primaryClass = {physics.comp-ph},
       adsurl = {https://ui.adsabs.harvard.edu/abs/2019ApJS..244...20Z}
}

@ARTICLE{lyon2004,
       author = {{Lyon}, J.~G. and {Fedder}, J.~A. and {Mobarry}, C.~M.},
        title = "{The Lyon-Fedder-Mobarry (LFM) global MHD magnetospheric simulation code}",
      journal = {Journal of Atmospheric and Solar-Terrestrial Physics},
         year = 2004,
        month = oct,
       volume = {66},
       number = {15-16},
        pages = {1333-1350},
          doi = {10.1016/j.jastp.2004.03.020},
       adsurl = {https://ui.adsabs.harvard.edu/abs/2004JASTP..66.1333L}
}

@ARTICLE{zhang2021,
       author = {{Zhang}, Binzheng and {Delamere}, Peter A. and {Yao}, Zhonghua and {Bonfond}, Bertrand and {Lin}, D. and {Sorathia}, Kareem A. and {Brambles}, Oliver J. and {Lotko}, William and {Garretson}, Jeff S. and {Merkin}, Viacheslav G. and {Grodent}, Denis and {Dunn}, William R. and {Lyon}, John G.},
        title = "{How Jupiter's unusual magnetospheric topology structures its aurora}",
      journal = {Science Advances},
         year = 2021,
        month = apr,
       volume = {7},
       number = {15},
        pages = {eabd1204},
          doi = {10.1126/sciadv.abd1204},
archivePrefix = {arXiv},
       eprint = {2006.14834},
 primaryClass = {physics.space-ph},
       adsurl = {https://ui.adsabs.harvard.edu/abs/2021SciA....7.1204Z}
}

@ARTICLE{chew1956,
       author = {{Chew}, G.~F. and {Goldberger}, M.~L. and {Low}, F.~E.},
        title = "{The Boltzmann Equation and the One-Fluid Hydromagnetic Equations in the Absence of Particle Collisions}",
      journal = {Proceedings of the Royal Society of London Series A},
         year = 1956,
        month = jul,
       volume = {236},
       number = {1204},
        pages = {112-118},
          doi = {10.1098/rspa.1956.0116},
       adsurl = {https://ui.adsabs.harvard.edu/abs/1956RSPSA.236..112C}
}

@ARTICLE{lyon1981,
       author = {{Lyon}, J.~G. and {Brecht}, S.~H. and {Huba}, J.~D. and {Fedder}, J.~A. and {Palmadesso}, P.~J.},
        title = "{Computer Simulation of a Geomagnetic Substorm}",
      journal = {\prl},
         year = 1981,
        month = apr,
       volume = {46},
       number = {15},
        pages = {1038-1041},
          doi = {10.1103/PhysRevLett.46.1038},
       adsurl = {https://ui.adsabs.harvard.edu/abs/1981PhRvL..46.1038L}
}

@ARTICLE{yee1966,
       author = {{Kane Yee}},
        title = "{Numerical solution of initial boundary value problems involving maxwell's equations in isotropic media}",
      journal = {IEEE Transactions on Antennas and Propagation},
         year = 1966,
        month = may,
       volume = {14},
       number = {3},
        pages = {302-307},
          doi = {10.1109/TAP.1966.1138693},
       adsurl = {https://ui.adsabs.harvard.edu/abs/1966ITAP...14..302Y}
}

@ARTICLE{mignone2014,
       author = {{Mignone}, A.},
        title = "{High-order conservative reconstruction schemes for finite volume methods in cylindrical and spherical coordinates}",
      journal = {Journal of Computational Physics},
         year = 2014,
        month = aug,
       volume = {270},
        pages = {784-814},
          doi = {10.1016/j.jcp.2014.04.001},
archivePrefix = {arXiv},
       eprint = {1404.0537},
 primaryClass = {physics.comp-ph},
       adsurl = {https://ui.adsabs.harvard.edu/abs/2014JCoPh.270..784M}
}

@ARTICLE{monchmeyer1989,
       author = {{Monchmeyer}, R. and {Muller}, E.},
        title = "{A Conservative Second-Order Difference Scheme for Curvilinear Coordinates - Part One - Assignment of Variables on a Staggered Grid}",
      journal = {\aap},
         year = 1989,
        month = jun,
       volume = {217},
        pages = {351},
       adsurl = {https://ui.adsabs.harvard.edu/abs/1989A&A...217..351M}
}

@ARTICLE{gottlieb2009,
       author = {{Gottlieb}, Sigal and {Ketcheson}, David I. and {Shu}, Chi-Wang},
        title = "{High Order Strong Stability Preserving Time Discretizations}",
      journal = {Journal of Scientific Computing},
         year = 2009,
        month = mar,
       volume = {38},
       number = {3},
        pages = {251-289},
          doi = {10.1007/s10915-008-9239-z},
       adsurl = {https://ui.adsabs.harvard.edu/abs/2009JSCom..38..251G}
}

@ARTICLE{evans1988,
       author = {{Evans}, Charles R. and {Hawley}, John F.},
        title = "{Simulation of Magnetohydrodynamic Flows: A Constrained Transport Model}",
      journal = {\apj},
         year = 1988,
        month = sep,
       volume = {332},
        pages = {659},
          doi = {10.1086/166684},
       adsurl = {https://ui.adsabs.harvard.edu/abs/1988ApJ...332..659E}
}

@ARTICLE{xu1999,
       author = {{Xu}, Kun},
        title = "{Gas-Kinetic Theory-Based Flux Splitting Method for Ideal Magnetohydrodynamics}",
      journal = {Journal of Computational Physics},
         year = 1999,
        month = aug,
       volume = {153},
       number = {2},
        pages = {334-352},
          doi = {10.1006/jcph.1999.6280},
       adsurl = {https://ui.adsabs.harvard.edu/abs/1999JCoPh.153..334X}
}

@article{rusanov1961,
  title={The calculation of the interaction of non-stationary shock waves with barriers},
  author={Rusanov, Viktor Vladimirovich},
  journal={Zhurnal Vychislitel'noi Matematiki i Matematicheskoi Fiziki},
  volume={1},
  number={2},
  pages={267--279},
  year={1961},
  publisher={Russian Academy of Sciences, Branch of Mathematical Sciences}
}

@ARTICLE{mignone2010,
       author = {{Mignone}, Andrea and {Tzeferacos}, Petros and {Bodo}, Gianluigi},
        title = "{High-order conservative finite difference GLM-MHD schemes for cell-centered MHD}",
      journal = {Journal of Computational Physics},
         year = 2010,
        month = aug,
       volume = {229},
       number = {17},
        pages = {5896-5920},
          doi = {10.1016/j.jcp.2010.04.013},
archivePrefix = {arXiv},
       eprint = {1001.2832},
 primaryClass = {astro-ph.HE},
       adsurl = {https://ui.adsabs.harvard.edu/abs/2010JCoPh.229.5896M}
}

@ARTICLE{leonard1991,
       author = {{Leonard}, B.~P. and {Niknafs}, H.~S.},
        title = "{Sharp monotonic resolution of discontinuities without clipping of narrow extrema}",
      journal = {Computers and Fluids},
         year = 1991,
        month = jan,
       volume = {19},
       number = {1},
        pages = {141-154},
       adsurl = {https://ui.adsabs.harvard.edu/abs/1991CF.....19..141L}
}

@ARTICLE{brambles2011magnetosphere,
       author = {{Brambles}, O.~J. and {Lotko}, W. and {Zhang}, B. and {Wiltberger}, M. and {Lyon}, J. and {Strangeway}, R.~J.},
        title = "{Magnetosphere Sawtooth Oscillations Induced by Ionospheric Outflow}",
      journal = {Science},
         year = 2011,
        month = jun,
       volume = {332},
       number = {6034},
        pages = {1183},
          doi = {10.1126/science.1202869},
       adsurl = {https://ui.adsabs.harvard.edu/abs/2011Sci...332.1183B}
}

@ARTICLE{balsara1999staggered,
       author = {{Balsara}, Dinshaw S. and {Spicer}, Daniel S.},
        title = "{A Staggered Mesh Algorithm Using High Order Godunov Fluxes to Ensure Solenoidal Magnetic Fields in Magnetohydrodynamic Simulations}",
      journal = {Journal of Computational Physics},
         year = 1999,
        month = mar,
       volume = {149},
       number = {2},
        pages = {270-292},
          doi = {10.1006/jcph.1998.6153},
       adsurl = {https://ui.adsabs.harvard.edu/abs/1999JCoPh.149..270B}
}

@Inbook{Shu1998,
author="Shu, Chi-Wang",

title="Essentially non-oscillatory and weighted essentially non-oscillatory schemes for hyperbolic conservation laws",
bookTitle="Advanced Numerical Approximation of Nonlinear Hyperbolic Equations: Lectures given at the 2nd Session of the Centro Internazionale Matematico Estivo (C.I.M.E.) held in Cetraro, Italy, June 23--28, 1997",
year="1998",
publisher="Springer Berlin Heidelberg",
address="Berlin, Heidelberg",
pages="325--432",
isbn="978-3-540-49804-9",
doi="10.1007/BFb0096355",
url="https://doi.org/10.1007/BFb0096355"
}

@ARTICLE{hain1987partial,
       author = {{Hain}, Klaus H.},
        title = "{The Partial Donor Cell Method}",
      journal = {Journal of Computational Physics},
         year = 1987,
        month = nov,
       volume = {73},
       number = {1},
        pages = {131-147},
          doi = {10.1016/0021-9991(87)90110-0},
       adsurl = {https://ui.adsabs.harvard.edu/abs/1987JCoPh..73..131H}
}

@ARTICLE{hau2002,
       author = {{Hau}, L.-N.},
        title = "{A note on the energy laws in gyrotropic plasmas}",
      journal = {Physics of Plasmas},
         year = 2002,
        month = jun,
       volume = {9},
       number = {6},
        pages = {2455-2457},
          doi = {10.1063/1.1476002},
       adsurl = {https://ui.adsabs.harvard.edu/abs/2002PhPl....9.2455H}
}

@ARTICLE{luo2023gas,
       author = {{Luo}, Hongyang and {Lyon}, John G. and {Zhang}, Binzheng},
        title = "{Gas kinetic schemes for solving the magnetohydrodynamic equations with pressure anisotropy}",
      journal = {Journal of Computational Physics},
         year = 2023,
        month = oct,
       volume = {490},
          eid = {112311},
        pages = {112311},
          doi = {10.1016/j.jcp.2023.112311},
archivePrefix = {arXiv},
       eprint = {2302.10922},
 primaryClass = {physics.plasm-ph},
       adsurl = {https://ui.adsabs.harvard.edu/abs/2023JCoPh.49012311L}
}

@ARTICLE{zhang2019conservative,
       author = {{Zhang}, Binzheng and {Sorathia}, Kareem A. and {Lyon}, John G. and {Merkin}, Viacheslav G. and {Wiltberger}, Michael},
        title = "{Conservative averaging-reconstruction techniques (Ring Average) for 3-D finite-volume MHD solvers with axis singularity}",
      journal = {Journal of Computational Physics},
         year = 2019,
        month = jan,
       volume = {376},
        pages = {276-294},
          doi = {10.1016/j.jcp.2018.08.020},
       adsurl = {https://ui.adsabs.harvard.edu/abs/2019JCoPh.376..276Z}
}

@ARTICLE{ji2006hydrodynamic,
       author = {{Ji}, Hantao and {Burin}, Michael and {Schartman}, Ethan and {Goodman}, Jeremy},
        title = "{Hydrodynamic turbulence cannot transport angular momentum effectively in astrophysical disks}",
      journal = {\nat},
         year = 2006,
        month = nov,
       volume = {444},
       number = {7117},
        pages = {343-346},
          doi = {10.1038/nature05323},
archivePrefix = {arXiv},
       eprint = {astro-ph/0611481},
 primaryClass = {astro-ph},
       adsurl = {https://ui.adsabs.harvard.edu/abs/2006Natur.444..343J}
}

@ARTICLE{gardiner2005unsplit,
       author = {{Gardiner}, Thomas A. and {Stone}, James M.},
        title = "{An unsplit Godunov method for ideal MHD via constrained transport}",
      journal = {Journal of Computational Physics},
         year = 2005,
        month = may,
       volume = {205},
       number = {2},
        pages = {509-539},
          doi = {10.1016/j.jcp.2004.11.016},
archivePrefix = {arXiv},
       eprint = {astro-ph/0501557},
 primaryClass = {astro-ph},
       adsurl = {https://ui.adsabs.harvard.edu/abs/2005JCoPh.205..509G}
}

@ARTICLE{orszag1979small,
       author = {{Orszag}, S.~A. and {Tang}, C.-M.},
        title = "{Small-scale structure of two-dimensional magnetohydrodynamic turbulence}",
      journal = {Journal of Fluid Mechanics},
         year = 1979,
        month = jan,
       volume = {90},
        pages = {129-143},
          doi = {10.1017/S002211207900210X},
       adsurl = {https://ui.adsabs.harvard.edu/abs/1979JFM....90..129O}
}

@ARTICLE{luo2025enhanced,
       author = {{Luo}, Hongyang and {Zhang}, Binzheng and {Lyon}, John G. and {Tian}, Jiaxing},
        title = "{Enhanced partial donor cell method for hyperbolic equations in orthogonal curvilinear coordinates}",
      journal = {Computer Physics Communications},
         year = 2025,
        month = nov,
       volume = {316},
          eid = {109808},
        pages = {109808},
          doi = {10.1016/j.cpc.2025.109808},
       adsurl = {https://ui.adsabs.harvard.edu/abs/2025CoPhC.31609808L}
}

@ARTICLE{stone1992zeus,
       author = {{Stone}, James M. and {Norman}, Michael L.},
        title = "{ZEUS-2D: A Radiation Magnetohydrodynamics Code for Astrophysical Flows in Two Space Dimensions. I. The Hydrodynamic Algorithms and Tests}",
      journal = {\apjs},
         year = 1992,
        month = jun,
       volume = {80},
        pages = {753},
          doi = {10.1086/191680},
       adsurl = {https://ui.adsabs.harvard.edu/abs/1992ApJS...80..753S}
}

@ARTICLE{dobler2006magnetic,
       author = {{Dobler}, Wolfgang and {Stix}, Michael and {Brandenburg}, Axel},
        title = "{Magnetic Field Generation in Fully Convective Rotating Spheres}",
      journal = {\apj},
         year = 2006,
        month = feb,
       volume = {638},
       number = {1},
        pages = {336-347},
          doi = {10.1086/498634},
archivePrefix = {arXiv},
       eprint = {astro-ph/0410645},
 primaryClass = {astro-ph},
       adsurl = {https://ui.adsabs.harvard.edu/abs/2006ApJ...638..336D}
}

@ARTICLE{dubey2008introduction,
       author = {{Dubey}, A. and {Reid}, L.~B. and {Fisher}, R.},
        title = "{Introduction to FLASH 3.0, with application to supersonic turbulence}",
      journal = {Physica Scripta Volume T},
         year = 2008,
        month = dec,
       volume = {132},
          eid = {014046},
        pages = {014046},
          doi = {10.1088/0031-8949/2008/T132/014046},
       adsurl = {https://ui.adsabs.harvard.edu/abs/2008PhST..132a4046D}
}

@ARTICLE{powell1999solution,
       author = {{Powell}, Kenneth G. and {Roe}, Philip L. and {Linde}, Timur J. and {Gombosi}, Tamas I. and {De Zeeuw}, Darren L.},
        title = "{A Solution-Adaptive Upwind Scheme for Ideal Magnetohydrodynamics}",
      journal = {Journal of Computational Physics},
         year = 1999,
        month = sep,
       volume = {154},
       number = {2},
        pages = {284-309},
          doi = {10.1006/jcph.1999.6299},
       adsurl = {https://ui.adsabs.harvard.edu/abs/1999JCoPh.154..284P}
}

@ARTICLE{ziegler2008nirvana,
       author = {{Ziegler}, U.},
        title = "{The NIRVANA code: Parallel computational MHD with adaptive mesh refinement}",
      journal = {Computer Physics Communications},
         year = 2008,
        month = aug,
       volume = {179},
       number = {4},
        pages = {227-244},
          doi = {10.1016/j.cpc.2008.02.017},
       adsurl = {https://ui.adsabs.harvard.edu/abs/2008CoPhC.179..227Z}
}

@ARTICLE{merkin2011disruption,
       author = {{Merkin}, V.~G. and {Lyon}, J.~G. and {McGregor}, S.~L. and {Pahud}, D.~M.},
        title = "{Disruption of a heliospheric current sheet fold}",
      journal = {\grl},
         year = 2011,
        month = jul,
       volume = {38},
       number = {14},
          eid = {L14107},
        pages = {L14107},
          doi = {10.1029/2011GL047822},
       adsurl = {https://ui.adsabs.harvard.edu/abs/2011GeoRL..3814107M}
}

@ARTICLE{kageyama2006note,
       author = {{Kageyama}, Akira and {Sugiyama}, Tooru and {Watanabe}, Kunihiko and {Sato}, Tetsuya},
        title = "{A note on the dipole coordinates}",
      journal = {Computers and Geosciences},
         year = 2006,
        month = mar,
       volume = {32},
       number = {2},
        pages = {265-269},
          doi = {10.1016/j.cageo.2005.06.006},
archivePrefix = {arXiv},
       eprint = {physics/0408133},
 primaryClass = {physics.geo-ph},
       adsurl = {https://ui.adsabs.harvard.edu/abs/2006CG.....32..265K}
}

@ARTICLE{zhang2024unified,
       author = {{Zhang}, B. and {Yao}, Z. and {Brambles}, O.~J. and {Delamere}, P.~A. and {Lotko}, W. and {Grodent}, D. and {Bonfond}, B. and {Chen}, J. and {Sorathia}, K.~A. and {Merkin}, V.~G. and {Lyon}, J.~G.},
        title = "{A unified framework for global auroral morphologies of different planets}",
      journal = {Nature Astronomy},
         year = 2024,
        month = aug,
       volume = {8},
        pages = {964-972},
          doi = {10.1038/s41550-024-02270-3},
       adsurl = {https://ui.adsabs.harvard.edu/abs/2024NatAs...8..964Z}
}

@ARTICLE{dang2023new,
       author = {{Dang}, Tong and {Zhang}, Binzheng and {Yan}, Maodong and {Lyon}, John and {Yao}, Zhonghua and {Xiao}, Sudong and {Zhang}, Tielong and {Lei}, Jiuhou},
        title = "{A New Tool for Understanding the Solar Wind-Venus Interaction: Three-dimensional Multifluid MHD Model}",
      journal = {\apj},
         year = 2023,
        month = mar,
       volume = {945},
       number = {2},
          eid = {91},
        pages = {91},
          doi = {10.3847/1538-4357/acba88},
       adsurl = {https://ui.adsabs.harvard.edu/abs/2023ApJ...945...91D}
}

@ARTICLE{lyon2025multifluid,
       author = {{Lyon}, J.~G. and {Merkin}, V.~G. and {Sorathia}, K.~A. and {Wiltberger}, M.~J.},
        title = "{Multifluid Equations for MHD}",
      journal = {Journal of Geophysical Research (Space Physics)},
         year = 2025,
        month = jun,
       volume = {130},
       number = {6},
          eid = {e2025JA033884},
        pages = {e2025JA033884},
          doi = {10.1029/2025JA033884},
       adsurl = {https://ui.adsabs.harvard.edu/abs/2025JGRA..13033884L}
}

@ARTICLE{mignone2011pluto,
       author = {{Mignone}, A. and {Zanni}, C. and {Tzeferacos}, P. and {van Straalen}, B. and {Colella}, P. and {Bodo}, G.},
        title = "{The PLUTO Code for Adaptive Mesh Computations in Astrophysical Fluid Dynamics}",
      journal = {\apjs},
         year = 2012,
        month = jan,
       volume = {198},
       number = {1},
          eid = {7},
        pages = {7},
          doi = {10.1088/0067-0049/198/1/7},
archivePrefix = {arXiv},
       eprint = {1110.0740},
 primaryClass = {astro-ph.HE},
       adsurl = {https://ui.adsabs.harvard.edu/abs/2012ApJS..198....7M}
}

@ARTICLE{colella2011high,
       author = {{Colella}, P. and {Dorr}, M.~R. and {Hittinger}, J.~A.~F. and {Martin}, D.~F.},
        title = "{High-order, finite-volume methods in mapped coordinates}",
      journal = {Journal of Computational Physics},
         year = 2011,
        month = apr,
       volume = {230},
       number = {8},
        pages = {2952-2976},
          doi = {10.1016/j.jcp.2010.12.044},
       adsurl = {https://ui.adsabs.harvard.edu/abs/2011JCoPh.230.2952C}
}

@ARTICLE{felker2018fourth,
       author = {{Felker}, Kyle Gerard and {Stone}, James M.},
        title = "{A fourth-order accurate finite volume method for ideal MHD via upwind constrained transport}",
      journal = {Journal of Computational Physics},
         year = 2018,
        month = dec,
       volume = {375},
        pages = {1365-1400},
          doi = {10.1016/j.jcp.2018.08.025},
archivePrefix = {arXiv},
       eprint = {1711.07439},
 primaryClass = {astro-ph.IM},
       adsurl = {https://ui.adsabs.harvard.edu/abs/2018JCoPh.375.1365F}
}

@ARTICLE{wang2017accurate,
       author = {{Wang}, Yulei and {Liu}, Jian and {Qin}, Hong and {Yu}, Zhi and {Yao}, Yicun},
        title = "{The accurate particle tracer code}",
      journal = {Computer Physics Communications},
         year = 2017,
        month = nov,
       volume = {220},
        pages = {212-229},
          doi = {10.1016/j.cpc.2017.07.009},
archivePrefix = {arXiv},
       eprint = {1609.07748},
 primaryClass = {physics.plasm-ph},
       adsurl = {https://ui.adsabs.harvard.edu/abs/2017CoPhC.220..212W}
}

@ARTICLE{merkin2010effects,
       author = {{Merkin}, V.~G. and {Lyon}, J.~G.},
        title = "{Effects of the low-latitude ionospheric boundary condition on the global magnetosphere}",
      journal = {Journal of Geophysical Research (Space Physics)},
         year = 2010,
        month = oct,
       volume = {115},
       number = {A10},
          eid = {A10202},
        pages = {A10202},
          doi = {10.1029/2010JA015461},
       adsurl = {https://ui.adsabs.harvard.edu/abs/2010JGRA..11510202M}
}

@ARTICLE{pembroke2012initial,
       author = {{Pembroke}, Asher and {Toffoletto}, Frank and {Sazykin}, Stanislav and {Wiltberger}, Michael and {Lyon}, John and {Merkin}, Viacheslav and {Schmitt}, Peter},
        title = "{Initial results from a dynamic coupled magnetosphere-ionosphere-ring current model}",
      journal = {Journal of Geophysical Research (Space Physics)},
         year = 2012,
        month = feb,
       volume = {117},
       number = {A2},
          eid = {A02211},
        pages = {A02211},
          doi = {10.1029/2011JA016979},
       adsurl = {https://ui.adsabs.harvard.edu/abs/2012JGRA..117.2211P}
}

@ARTICLE{sorathia2020ballooning,
       author = {{Sorathia}, K.~A. and {Merkin}, V.~G. and {Panov}, E.~V. and {Zhang}, B. and {Lyon}, J.~G. and {Garretson}, J. and {Ukhorskiy}, A.~Y. and {Ohtani}, S. and {Sitnov}, M. and {Wiltberger}, M.},
        title = "{Ballooning-Interchange Instability in the Near-Earth Plasma Sheet and Auroral Beads: Global Magnetospheric Modeling at the Limit of the MHD Approximation}",
      journal = {\grl},
         year = 2020,
        month = jul,
       volume = {47},
       number = {14},
          eid = {e88227},
        pages = {e88227},
          doi = {10.1029/2020GL088227},
       adsurl = {https://ui.adsabs.harvard.edu/abs/2020GeoRL..4788227S}
}

@ARTICLE{provornikova2024mhd,
       author = {{Provornikova}, Elena and {Merkin}, Viacheslav G. and {Vourlidas}, Angelos and {Malanushenko}, Anna and {Gibson}, Sarah E. and {Winter}, Eric and {Arge}, Charles N.},
        title = "{MHD Modeling of a Geoeffective Interplanetary Coronal Mass Ejection with the Magnetic Topology Informed by In Situ Observations}",
      journal = {\apj},
         year = 2024,
        month = dec,
       volume = {977},
       number = {1},
          eid = {106},
        pages = {106},
          doi = {10.3847/1538-4357/ad83b1},
archivePrefix = {arXiv},
       eprint = {2405.13069},
 primaryClass = {astro-ph.SR},
       adsurl = {https://ui.adsabs.harvard.edu/abs/2024ApJ...977..106P}
}
\bibliographystyle{aasjournal}

\end{document}